\documentclass[11pt]{amsart}
\usepackage{geometry}                
\usepackage{subfig}
\usepackage{graphicx}
\usepackage{amssymb, bbm,mathrsfs,array}
\usepackage{epstopdf}
\usepackage{multirow}
\usepackage{hyperref}
\usepackage[foot]{amsaddr}

\theoremstyle{remark}

\newcommand{\bo}[1]{\boldsymbol{#1}}
\DeclareMathOperator{\diag}{diag}

\title{A New Twisting Somersault - 513XD}
\author{William Tong and Holger R.~Dullin}
\address{The University of Sydney, School of Mathematics and Statistics}
\email{william.tong@sydney.edu.au, holger.dullin@sydney.edu.au}

\begin{document}
\maketitle


\begin{abstract}
We present the mathematical framework of an athlete modelled as a system of coupled rigid bodies to simulate platform and springboard diving. Euler's equations of motion are generalised to non-rigid bodies, and are then used to innovate a new dive sequence that in principle can be performed by real world athletes. We begin by assuming shape changes are instantaneous so that the equations of motion simplify enough to be solved analytically, and then use this insight to present a new dive (513XD) consisting of 1.5 somersaults and 5 twists using realistic shape changes. Finally, we demonstrate the phenomenon of converting pure somersaulting motion into pure twisting motion by using a sequence of impulsive shape changes, which may have applications in other fields such as space aeronautics.
\end{abstract}


\section{Introduction}
The twisting somersault is a visually stunning acrobatic manoeuvre featured in numerous Olympic sports. Within diving alone there is a wide selection of books available to assist athletes and coaches of all skill levels, e.g.\! \cite{batterman, Fairbanks, Huber, Moriarty, obrien, Still}. Coaches are now seeking to better understand the biomechanics behind the aerial manoeuvres associated with a good dive to equip their athletes with a leading edge in competition. The purpose of this paper is to use the newly established mathematical framework developed in \cite{Rotor, TwistSom, WTongthesis} to present the innovative 513XD\footnote{The FINA diving code 513XD has the following interpretation: the initial 51 stands for "twisting forward", the following digits are the number of half-somersaults (3) and the number of half-twists (10=X), and the final letter D stands for "free" position.}
dive consisting of 1.5 somersaults and 5 twists that in principle can be performed by world class athletes. In doing so, we develop a mathematical understanding of the effects of shape change on the dynamics of the system. 

The physics behind the twisting somersault was first correctly described in detail by Frohlich in \cite{Frohlich}, where he explains how a diver taking off in pure somersaulting motion can utilise shape change to initiate twist mid-flight. Since then, Yeadon has extensively analysed aerial movement of the human body in \cite{yeadon90a, yeadon90b, yeadon90c, yeadon90d} and the biomechanics of the twisting somersault in \cite{yeadon93a, yeadon93b, yeadon93c, yeadon93d}. The simplest mechanism for producing the effect of twisting somersault is found in \cite{Rotor}, which uses a rotor instead of arms to initiate and terminate twist. As the shape is not physically changing, the system is simple enough so that an analytical formula can be established connecting the number of somersaults, twists and airborne time. The model in \cite{TwistSom} incorporates shape change to resemble real dives, however simplicity is kept by fixing nearly all segments relative to the torso, with the exception of the left shoulder joint. In this paper we build on the results established in \cite{TwistSom} and allow movement in both arms with the purpose of presenting the innovative 513XD dive. To date, no athlete has attempted a 513XD dive in competition, nor has the International Swimming Federation (FINA) even assigned a degree-of-difficulty to it. By simulating the 513XD dive we hope to provide coaches and athletes with insight and motivation so that the dive may one day be executed in competition.

An original simplified version of the 513XD dive was discovered by the 2nd author in November 2013, while on sabbatical in Boulder, Colorado.
The details of the dive and important improvements were worked out in the PhD thesis \cite{WTongthesis} of the first author.
The mathematical core of our analysis is the use of geometric mechanics to achieve a conceptual understanding of successful dives.
This allows for the description of the motion in a symmetry reduced space (i.e.~the body frame), from which the 
motion in full phase space can be reconstructed using a formula involving a so called geometric phase and a dynamic phase.
Even for rigid body motion such a description is relatively new \cite{Montgomery90}, and the extension to non-rigid bodies 
\cite{Cabrera07} is quite recent. In \cite{TwistSom} we have extended their formulas to remove a $\bmod 2\pi$ operation;
this extension is essential for application to diving. 
A related but different extension has been introduced earlier in \cite{Cushman05}.
Using these modern tools from geometric mechanics we are thus able to obtain a full 
understanding of twisting somersault, and as a result can propose a
dive that has never been performed as of the end of 2016.

The structure of the paper is as follows: in section \ref{sec:model} we introduce the model of the athlete consisting of three segments, and establish the overall tensor of inertia $I$ and momentum shift $\bo{A}$ of the system. In section \ref{sec:eqsystem} we develop the mathematical framework by deriving the equations of motion and presenting the solutions to the dynamics of rigid body motion. In section \ref{sec:fastkick} we use impulsive shape changes to construct the faster twisting somersault, and compare our findings to the twisting somersault described in \cite{TwistSom}. In section \ref{sec:513XD} we present the 513XD dive consisting of 1.5 somersaults and 5 twists using realistic shape changes. We show that this dive can in principle be executed by real world athletes, and if successfully performed in competition would revolutionise the sport of diving. Finally, in section \ref{sec:som2twi} we demonstrate it is possible for an athlete to take-off in pure somersaulting motion and use a sequence of impulsive shape changes to enter a state of pure twist. This shows the athlete can continually apply shape changes to speed up twist in the twisting somersault, limited only when pure twisting motion is achieved.

\section{Model of the athlete}\label{sec:model}
In the literature there are several mathematical models of the human body varying in complexity, ranging from Hanavan's model \cite{hanavan} comprised of simple geometric solids such as ellipsoids and truncated cones, to Jensen's model \cite{Jensen76} that involves stacking thin elliptical disks atop one other for better approximation of each segment. However, when examining the dynamics only the moments of inertia, centre of mass and joint positions of each segment come into the play, leaving all of the complicated geometries hidden within the system. For this reason we are not overly concerned with the geometry and thus base our three segment model off Frohlich's \cite{Frohlich}, where his model parameters are presented in Appendix \ref{app:modelparameters}. The mass $m_i$ of each segment $B_i$ is evaluated by summing up the masses of the subsegments given by Frohlich's \cite{Frohlich} model, and we denote $M$ as the total mass of the athlete. From here on we will use index $i \in\{b,l,r\}$ to refer to body segment $B_i$.

\begin{figure}[t]
\centering
\subfloat[The body segments $B_i$, local body frames $\mathcal{F}_i$ and overall body frame $\mathcal{F}_C$ (dashed axes).]{\includegraphics[width=7.29cm]{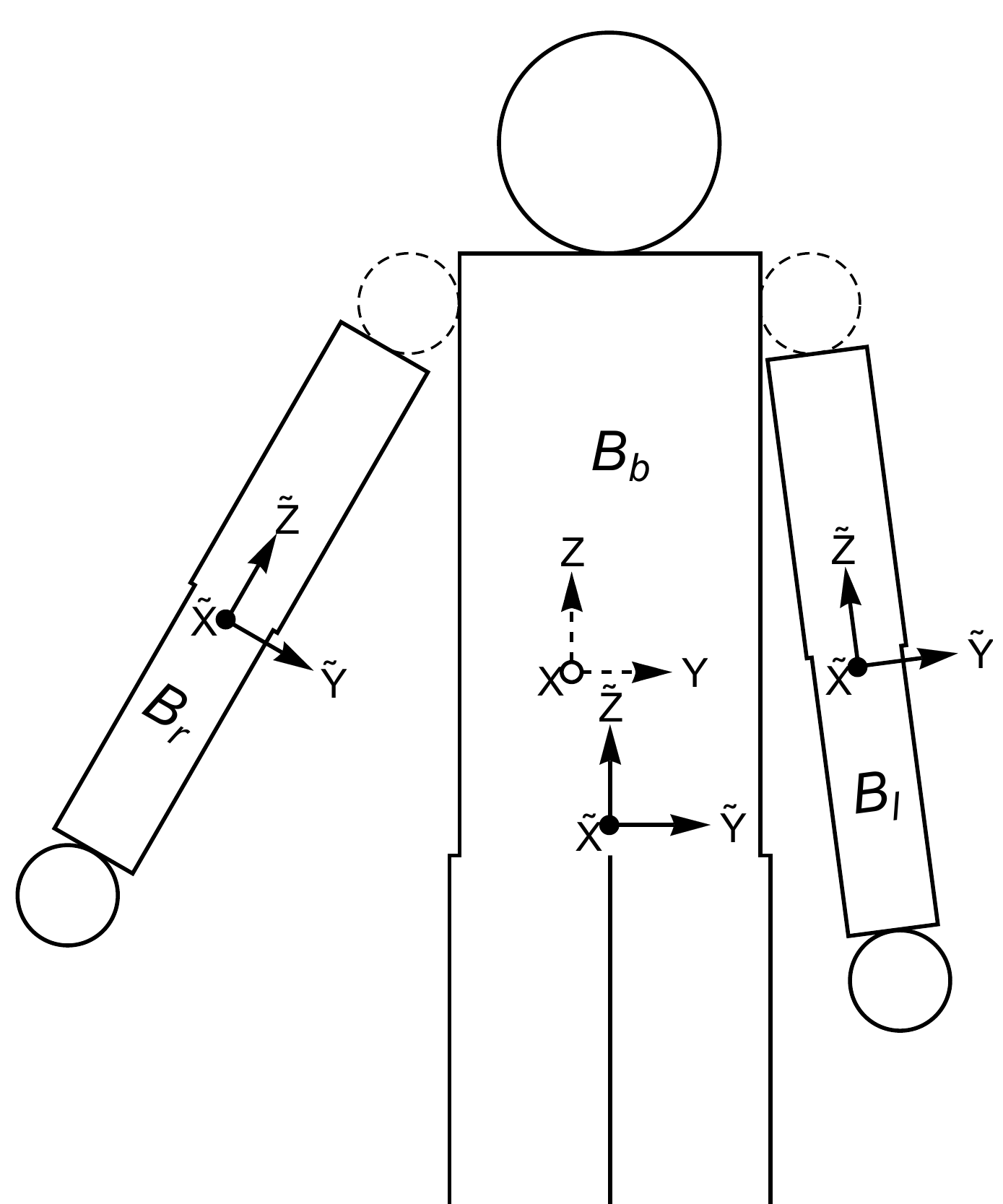}\label{subfig:modela}}
\subfloat[The centre of mass vectors $\bo{C}_i$ (solid vectors) and joint vectors $\bo{\tilde{J}}_i^j$ (dashed vectors).]{\includegraphics[width=7.29cm]{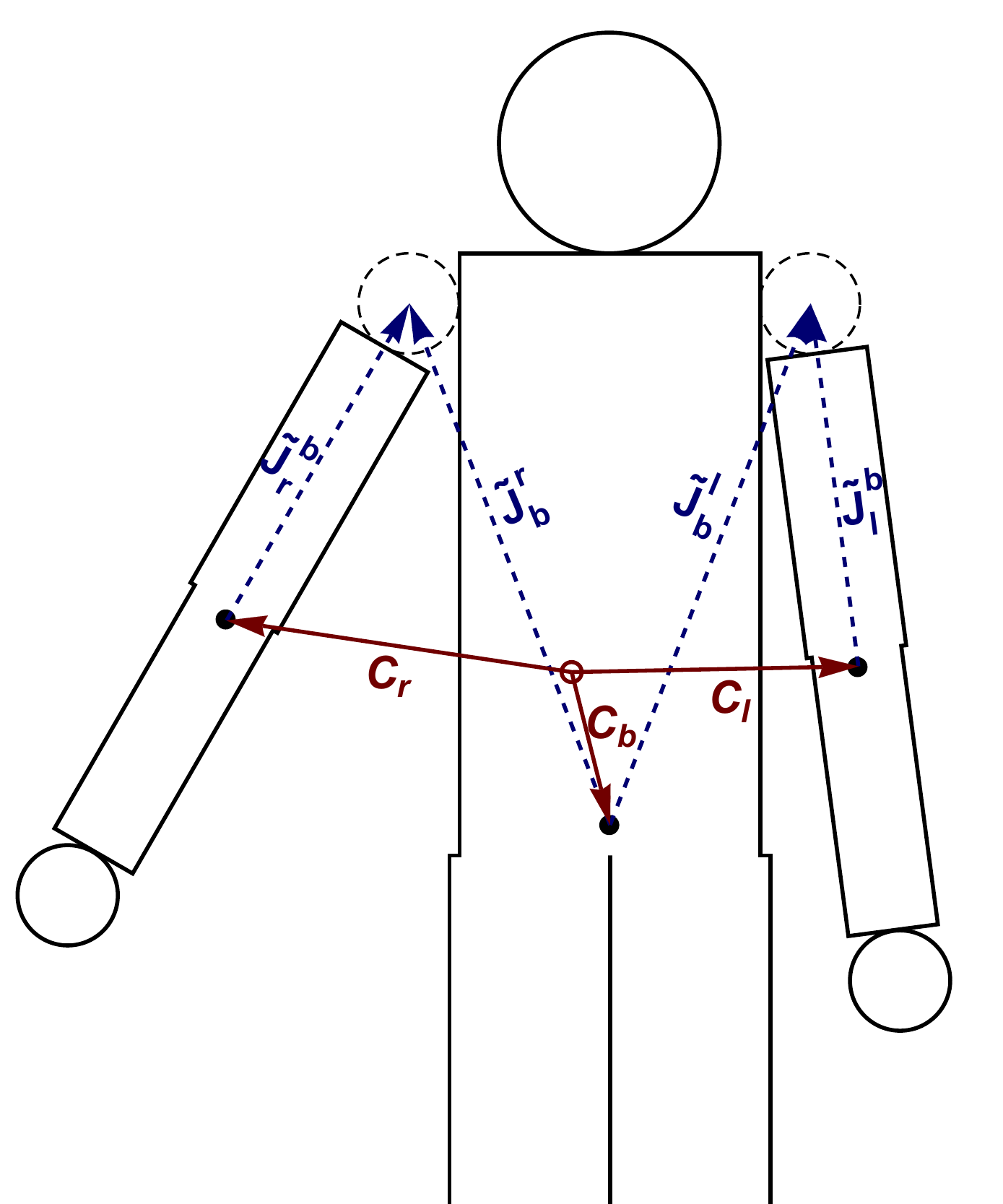}\label{subfig:modelb}}
\caption{Visualisation of body segments, body frames, centre of mass and joint vectors. We note that the origin of $\mathcal{F}_C$ lies close to $\mathcal{F}_b$ in reality, so for illustration purposes we have deliberately moved the origin of $\mathcal{F}_C$ to clearly display the local body frames and vectors. The lower legs and feet have also been omitted for this reason.}\label{fig:model}
\end{figure}

For each $B_i$ a local body frame $\mathcal{F}_i$ is attached with its origin coinciding with the centre of mass and coordinate axes pointing in the direction of the principal moments of inertia, where the tensor of inertia $\tilde{I}_i$ is computed using the parallel axis theorem. The joint vector $\bo{\tilde{J}}_i^j$ is written in the local body frame $\mathcal{F}_i$ so that it is constant, and the geometric interpretation is the position vector from the centre of mass of $B_i$ to the joint location that connects to $B_j$. We illustrate body segments $B_i$, frames $\mathcal{F}_i$, joint vectors $\bo{\tilde{J}}_i^j$ and centre of mass vectors $\bo{C}_i$ in Figure \ref{fig:model}, and provide the numerical values of the collection of $\tilde{I}_i$ and $\bo{\tilde{J}}_i^j$ for our specific model in Appendix \ref{app:modelparameters}.

The coupled rigid body system has a spatial frame $\mathcal{F}_S$ and overall body frame $\mathcal{F}_C$, where both frames are chosen to have their origin co-moving with the overall centre of mass of the system. Now the coordinate axes of $\mathcal{F}_S$ are fixed in direction, and without loss of generality we choose to align the axes of $\mathcal{F}_C$ with the torso specified by $\mathcal{F}_b$. We will use lowercase letters to denote quantities in $\mathcal{F}_S$, uppercase letters in $\mathcal{F}_C$ and uppercase letters with tilde in $\mathcal{F}_i$. Let $\bo{V}$ be any arbitrary vector observed in $\mathcal{F}_C$, so that the very same vector in $\mathcal{F}_S$ is given by $\bo{v}$. The transformation between the two frames is given by $\bo{v}=R\bo{V}$, where $R=R(t)$ is the rotation matrix that specifies the orientation of the system. The shape of the system is specified by the collection of rotation matrices $\{R_i\}$, where based on our choice of $\mathcal{F}_C$ we have $R_b=\mathbbm{1}$. To express an arbitrary vector $\bo{\tilde{V}}_i$ written in $\mathcal{F}_i$ to the overall body frame $\mathcal{F}_C$, we use the transformation $\bo{V}=\bo{C}_i+R_i \bo{\tilde{V}}_i$. 

We now want to express $\bo{C}_i$ in terms of the collection of geometry $\{\bo{\tilde{J}}_i^j\}$ and shape $\{R_i\}$. Provided that $B_i$ and $B_j$ are two connected bodies we have
\begin{equation}
\bo{C}_i+R_i \bo{\tilde{J}}_i^j = \bo{C}_j + R_j \bo{\tilde{J}}_j^i,\label{eq:jointcond}
\end{equation}
which can be verified with Figure \ref{subfig:modelb}. By construction $\displaystyle\sum_{i} m_i \bo{C}_i=\bo{0}$, so using \eqref{eq:jointcond} to solve for a reference centre we get
\begin{equation}
\bo{C}_b = \frac{1}{M}\sum_{i\in\{l,r\}}m_i(R_i\bo{\tilde{J}}_i^b-\bo{\tilde{J}}_b^i),\label{eq:Cb}
\end{equation}
which can be substituted back in \eqref{eq:jointcond} to obtain
\begin{equation}
\bo{C}_i = \bo{C}_b + \bo{\tilde{J}}_b^i-R_i \bo{\tilde{J}}_i^b.
\end{equation}
Note in the case of $i=b$ we have $R_b=\mathbbm{1}$, so we simply retrieve \eqref{eq:Cb} as expected.

As derived in \cite{TwistSom}, for a system of coupled rigid bodies we have
\begin{align}
I &= \sum_{i}\left(R_i \tilde{I}_i R_i^t+m_i[|\bo{C}_i|^2\mathbbm{1}-\bo{C}_i\bo{C}_i^t]\right)\label{eq:I}\\
\bo{A} &= \sum_{i}\left(m_i \bo{C}_i\times\bo{\dot{C}}_i+R_i \tilde{I}_i\bo{\Omega}_i\right),\label{eq:A}
\end{align}
where $I=I(t)$ is the tensor of inertia of the system and $\bo{A}=\bo{A}(t)$ is the momentum shift generated by shape change. The angular velocity $\bo{\Omega}_i$ for $B_i$ is relative with respect to $B_b$, and is defined such that $R_i^t \dot{R}_i \bo{V} = \hat{\bo{\Omega}}_i\bo{V}$ for some arbitrary vector $\bo{V}$. 

If we restrict the arm movement to be about the abduction-adduction plane of motion as shown in Figure \ref{fig:model}, then the shape of the athlete can be completely specified with just two angles $(\alpha_l,\alpha_r)\in[0,\pi]^2$. We let $(\alpha_l,\alpha_r)=(0,0)$ correspond to the anatomical neutral position where the arms are down by the side, and $(\alpha_l,\alpha_r)=(\pi,\pi)$ be the layout position where both arms are pointing straight up. In terms of rotation matrices we have $R_l(\alpha_l)=\mathcal{R}_x(\alpha_l)$ and $R_r(\alpha_r)=\mathcal{R}_x(-\alpha_r)$, where $\mathcal{R}_x$ is the elementary rotation matrix about the $x$-axis and the minus sign accounts for the opposite direction of right arm rotation when compared to the left arm. 
For simplicity we will introduce a two letter arm code to describe the actions of the left and right arms, respectively. When the arm is stationary relative to the torso we use U for the up position and D for the down position, or if the arm is in motion we use H when it is being raised and L when it is being lowered. E.g.\! the layout position is denoted by UU as both arms are pointing straight up, and the anatomical neutral position is denoted by DD as both arms are down by the side.
Evaluating the tensor of inertia $I$ with these shape change restrictions simplifies it to the form 
\begin{equation}
I = \left(\begin{array}{ccc}
I_{xx} & 0      & 0\\
0      & I_{yy} & I_{yz}\\
0      & I_{yz} & I_{zz}
\end{array}\right),\label{eq:I3}
\end{equation} 
and similarly the momentum shift $\bo{A}$ reduces to 
\begin{equation}
\bo{A}=\big(A_l \dot{\alpha}_l+A_r \dot{\alpha}_r,0,0\big)^t.\label{eq:A3}
\end{equation}
The components of $I$ and $\bo{A}$ are explicitly listed in Appendix \ref{app:components}. In general the tensor of inertia $I$ is non-diagonal, but there is always a  coordinate transformation 
\begin{equation}
J = \diag{(J_x, J_y, J_z)} =  R^{-1}_p I R_p\label{eq:J}
\end{equation}
with some rotation matrix $R_p$ that makes $J$ diagonal in some alternate body frame $\mathcal{F}_P$.
In the block-diagonal case of \eqref{eq:I3} we have 
\begin{equation}
R_p=\mathcal{R}_x(p) \qquad\text{where}\qquad p=\frac{1}{2}\arctan{\left(\frac{2 I_{yz}}{I_{yy}-I_{zz}}\right)}.\label{eq:p}
\end{equation}
We denote $I_s=\diag(I_{s,x},I_{s,y},I_{s,z})$ as the tensor of inertia for the layout position given by shape UU, and $J_t=\diag(J_{t,x},J_{t,y}J_{t,z})$ as the diagonalised tensor of inertia for the twist position given by shape DU and UD. We want to emphasise that $J_t$ is written in $\mathcal{F}_P$, and the transformation \eqref{eq:J} can be used to rewrite tensors in $\mathcal{F}_C$. The numerical values of $I_s$ and $J_t$ are specified in Appendix \ref{app:modelparameters}.

\section{Equations of motion}\label{sec:eqsystem}
\begin{figure}[t]
\centering
\hspace{-9mm}\subfloat[Cartwheel $\bo{L}=(l,0,0)^t$]{\includegraphics[width=5.35cm]{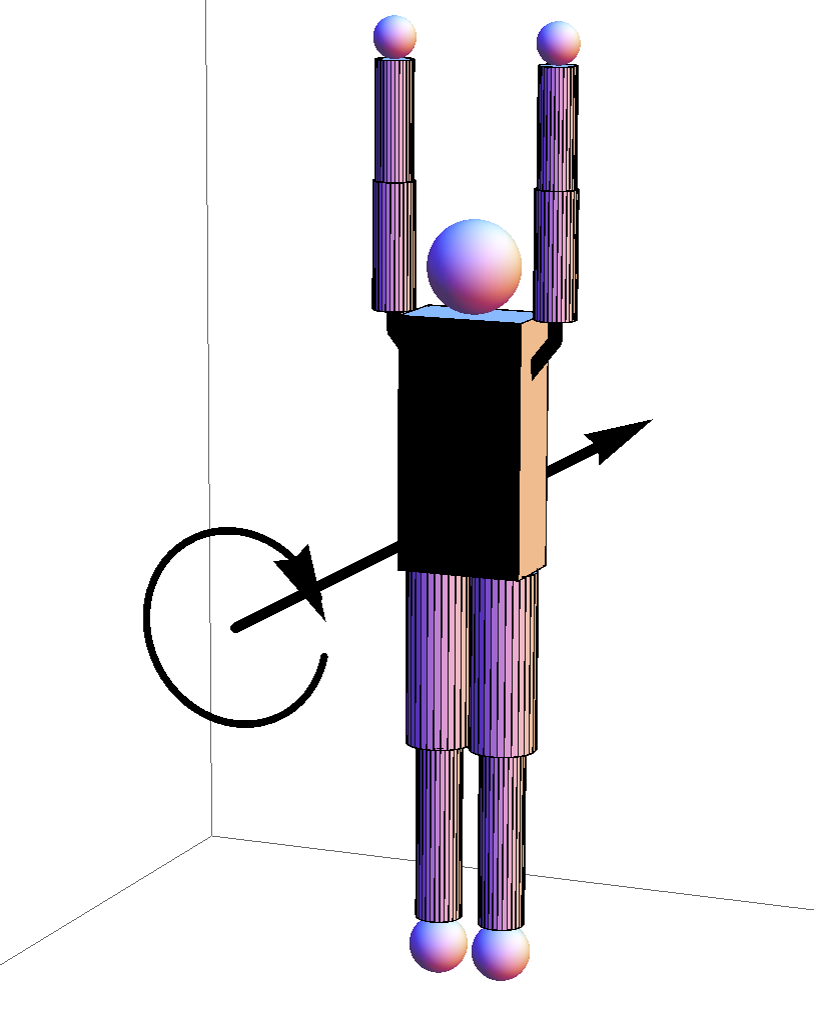}\label{subfig:cartwheel}}
\subfloat[Somersault $\bo{L}=(0,l,0)^t$]{\includegraphics[width=5.35cm]{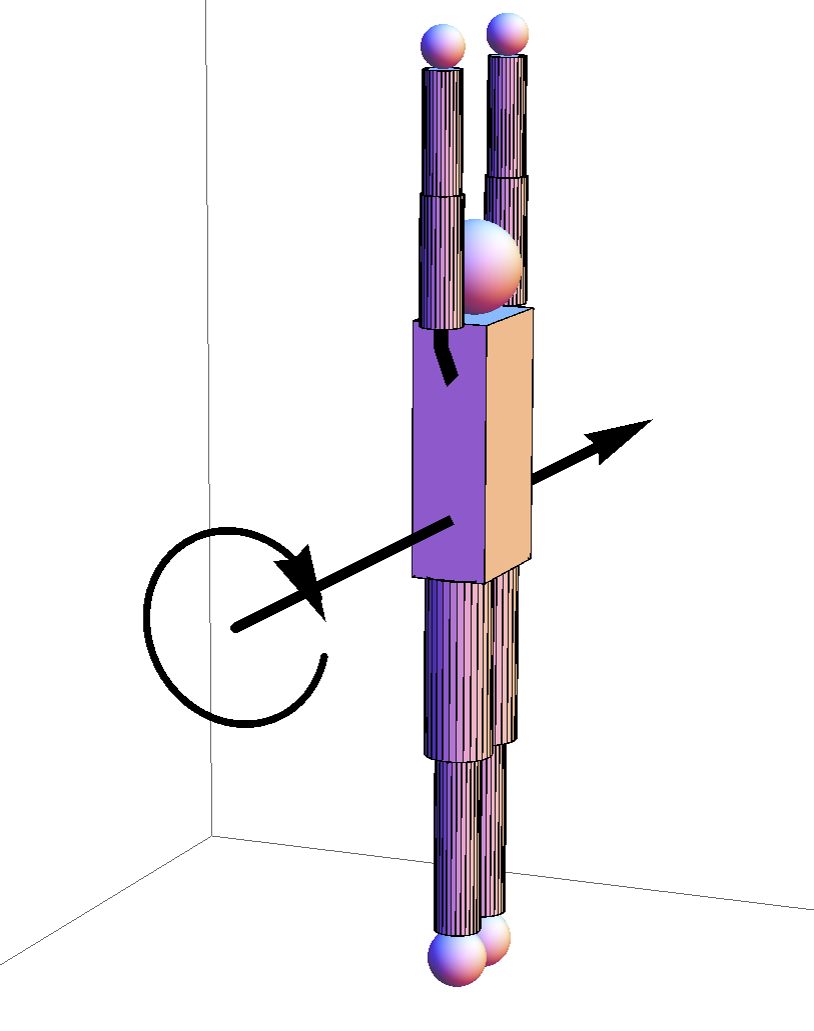}\label{subfig:somersault}}
\subfloat[Twist $\bo{L}=(0,0,l)^t$]{\includegraphics[width=5.3cm]{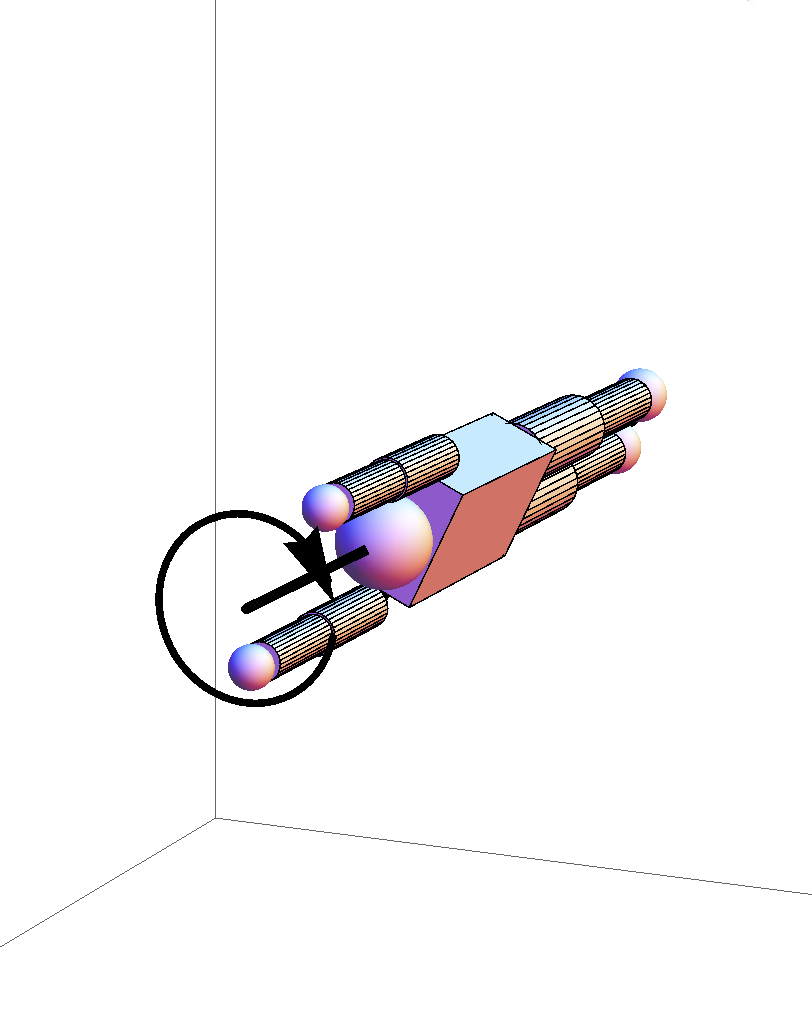}\label{subfig:twist}}
\caption{Each diagram above corresponds to steady rotations, where the direction can be either clockwise or counterclockwise resulting in a total of six equilibria. To distinguish between the front and back of the athlete's torso, the back has been shaded black.}\label{fig:equilibria}
\end{figure}

As shown in \cite{TwistSom} the angular momentum vector $\bo{L}$ of a coupled rigid body system can be expressed as
\begin{equation}
\bo{L}=I\bo{\Omega}+\bo{A},
\end{equation}
where $I$ is the tensor of inertia, $\bo{A}$ is the momentum shift, $\bo{\Omega}$ is the angular velocity and all quantities are viewed from $\mathcal{F}_C$. In the absence of shape change $I$ is constant and $\bo{A}=\bo{0}$ because it is linear in the shape velocities, thus we recover the well known formula $\bo{L}=I\bo{\Omega}$ for rigid body dynamics. To derive the equations of motion for coupled rigid bodies we take the time derivative of $\bo{l}=R\bo{L}$, which gives $\bo{\dot{l}}=\dot{R}\bo{L}+R\bo{\dot{L}}=\bo{0}$ and use $R^t \dot{R}\bo{V}=\hat{\bo{\Omega}}V = \bo{\Omega}\times\bo{V}$ to express the result as 
\begin{equation}
\bo{\dot{L}}= \bo{L}\times\bo{\Omega}=\bo{L}\times I^{-1}(\bo{L}-\bo{A}).\label{eq:eom}
\end{equation}
When there is no shape change (which implies that $\bo{A}=0$ and $I = const$)
there are the classical six equilibria of steady rotations, which are illustrated in Figure \ref{fig:equilibria} for symmetric shape UU with spatial angular momentum vector $\bo{l}=(0,l,0)^t$. For asymmetric shape $(\alpha_l,\alpha_r)$ the equilibrium points are rotated by $R_p$ given by \eqref{eq:p}, which is due to the diagonalisation of $I$ shown in \eqref{eq:J}. In our model of the athlete the diagonalised tensor of inertia $J=\diag{(J_x, J_y, J_z)}$ has components $J_x>J_y>J_z$ for all $(\alpha_l,\alpha_r)\in[0,\pi]^2$.
It is a classical result that rotation about the axis of the intermediate moment of inertia is 
unstable. Thus cartwheeling and twisting motions are stable while somersaulting motion is unstable irrespective of the arm positions.

In the absence of external forces and without shape change the orbit $\bo{L}$ is governed by conservation laws, where the total energy $E$ and spatial angular momentum vector $\bo{l}$ is constant. As the rotation matrix $R_p$ can be used to diagonalise an arbitrary tensor $I$, we will assume without loss of generality that $\mathcal{F}_C$ is aligned with the principal moments of inertia and rotates by \eqref{eq:p} whenever necessary. This enables us to write $\bo{L}=(L_x,L_y,L_z)^t$ and $I=\diag(I_x,I_y,I_z)$ to avoid introducing new symbols for these quantities in another body frame. The angular kinetic energy $E=\frac{1}{2}\bo{\Omega}^t I \bo{\Omega}$ defines a surface known as Poinsot's ellipsoid, and rewriting this ellipsoid in terms of $\bo{L}$ produces what we refer to as the energy-inertia ellipsoid
\begin{equation}
E=\frac{L_x^2}{2I_x}+\frac{L_y^2}{2I_y}+\frac{L_z^2}{2I_z}.\label{eq:conserveE}
\end{equation}
As the angular momentum $\bo{l}$ is constant in $\mathcal{F}_S$ the length is preserved in $\mathcal{F}_C$, which means $\bo{L}$ must lie on the $\bo{L}$-sphere with equation
\begin{equation}
l^2 = L_x^2+L_y^2+L_z^2.\label{eq:conserveL}
\end{equation} 
Thus the trajectory of $\bo{L}$ must lie on the intersection between the energy-inertia ellipsoid \eqref{eq:conserveE} and $\bo{L}$-sphere \eqref{eq:conserveL}, which in general forms two closed curves.
The time evolution on these curves can be expressed in terms of Jacobi elliptic functions, see, e.g.\! \cite{cushman91, mech}, which we write as
\begin{equation}
\bo{\mathcal{L}}(t; E, I, c)= \big(L_x(t),L_y(t),L_z(t)\big)^t\label{eq:Lorbit}
\end{equation}
where the parameters on the right hand side are suppressed. Specifically, when $l^2 I_y^{-1}<2E<l^2 I_z^{-1}$ the components are
\begin{align}
L_x &= s\sqrt{\frac{I_x (l^2-2E I_z)}{I_x-I_z}}\operatorname{cn}{(\tau,k^2)} & L_y &= \sqrt{\frac{I_y (l^2-2E I_z)}{I_y-I_z}}\operatorname{sn}{(\tau,k^2)}
\label{eq:Lsol}\\
L_z &= -s\sqrt{\frac{I_z (2E I_x - l^2)}{I_x-I_z}}\operatorname{dn}{(\tau,k^2)}\nonumber
\end{align}
with
\begin{align}
\tau &= \sqrt{\frac{(I_y-I_z)(2E I_x-l^2)}{I_x I_y I_z}}\left(t+c\right) & 
k^2 &= \frac{(I_x-I_y)(l^2-2E I_z)}{(I_y-I_z)(2E I_x-l^2)},\label{eq:tk}
\end{align}
and the two constants $c$ (phase shift) and $s$ (direction that is either $\pm 1$)
 that appear are chosen to satisfy the initial conditions. As we will only be using counterclockwise twists in our computations, $s = +1$ always, and thus we omit this constant in the parameters of \eqref{eq:Lorbit}. It is important to note the minus sign in \eqref{eq:Lsol} for $L_z$ occurs because $I_x>I_y>I_z$ for the diver, and that had the inequalities been reversed, i.e.\! $I_x<I_y<I_z$, there would be no minus sign. As the Jacobi elliptic functions are periodic in $\tau$ with period $4K(k^2)=4\displaystyle\int_0^\frac{\pi}{2}\frac{du}{\sqrt{1-k^2\sin^2{u}}}$, the period of the orbit is 
\begin{figure}[b]
\includegraphics[width=9.2cm]{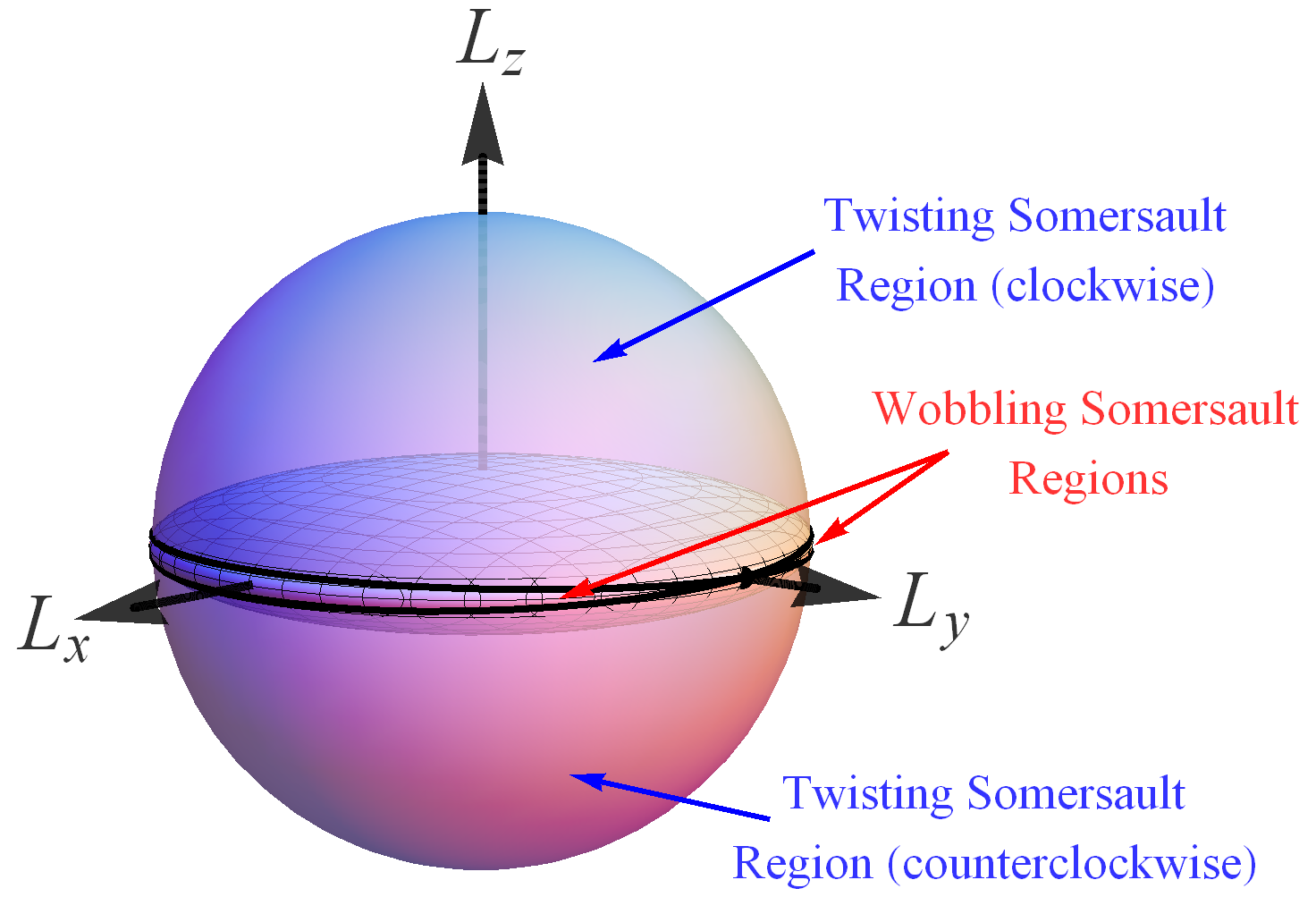}
\caption{Intersection of the energy-inertia ellipsoid with $\bo{L}$-sphere when $2E = l^2 I_y^{-1}$ results in four trajectories emerging from the poles $(0,\pm l,0)$. The separatrices divide the dynamics into the twisting somersault and wobbling somersault regions as indicated.}
\label{fig:sphereY}
\end{figure}
\begin{equation}
T(E, I) = 4K(k^2)\sqrt{\frac{I_x I_y I_z}{(I_y-I_z)(2EI_x-l^2)}}\label{eq:T}
\end{equation}
where $K(k^2)$ is the complete elliptic integral of the first kind. When $l^2 I_x^{-1}<2E<l^2 I_y^{-1}$, the results presented in \eqref{eq:Lsol}, \eqref{eq:tk} and \eqref{eq:T} need to be modified by swapping $L_x\leftrightarrow L_z$ and $I_x\leftrightarrow I_z$. In the limiting case when $2E\rightarrow l^2 I_x^{-1}$ the energy becomes minimal and the two closed curves shrink to the points given by $(\pm l,0,0)$. Similarly, when $2E\rightarrow l^2 I_z^{-1}$ the energy becomes maximal and the curves shrink to the points at $(0,0,\pm l)$. When $2E = l^2 I_y^{-1}$ the Jacobi elliptic functions reduce to hyperbolic functions given by  
\begin{align}
L_x &= s_1 l\sqrt{\frac{I_x(I_y-I_z)}{I_y(I_x-I_z)}}\operatorname{sech}{\tau} &
L_y &= s_2 l\tanh{\tau} &
L_z &= -s_1 s_2 l\sqrt{\frac{I_z (I_x-I_y)}{I_y(I_x-I_z)}}\operatorname{sech}{\tau},
\end{align}
where the two signs $s_1$ and $s_2$ distinguish the four trajectories resulting from the intersection of the $\bo{L}$-sphere and energy-inertia ellipsoid shown in Figure \ref{fig:sphereY}. The sign $s_2$ determines whether we are on the stable or unstable manifold of the unstable equilibrium points, and $s_1$ distinguishes which of the two branches of the manifold we are on. 
From the unstable equilibrium points it is easy to move into the twisting region by a small kick increasing $L_z$, which will 
initiate twisting motion travelling all the way to the other side of the sphere before returning. 
\begin{figure}[b]
\includegraphics[width=7.4cm]{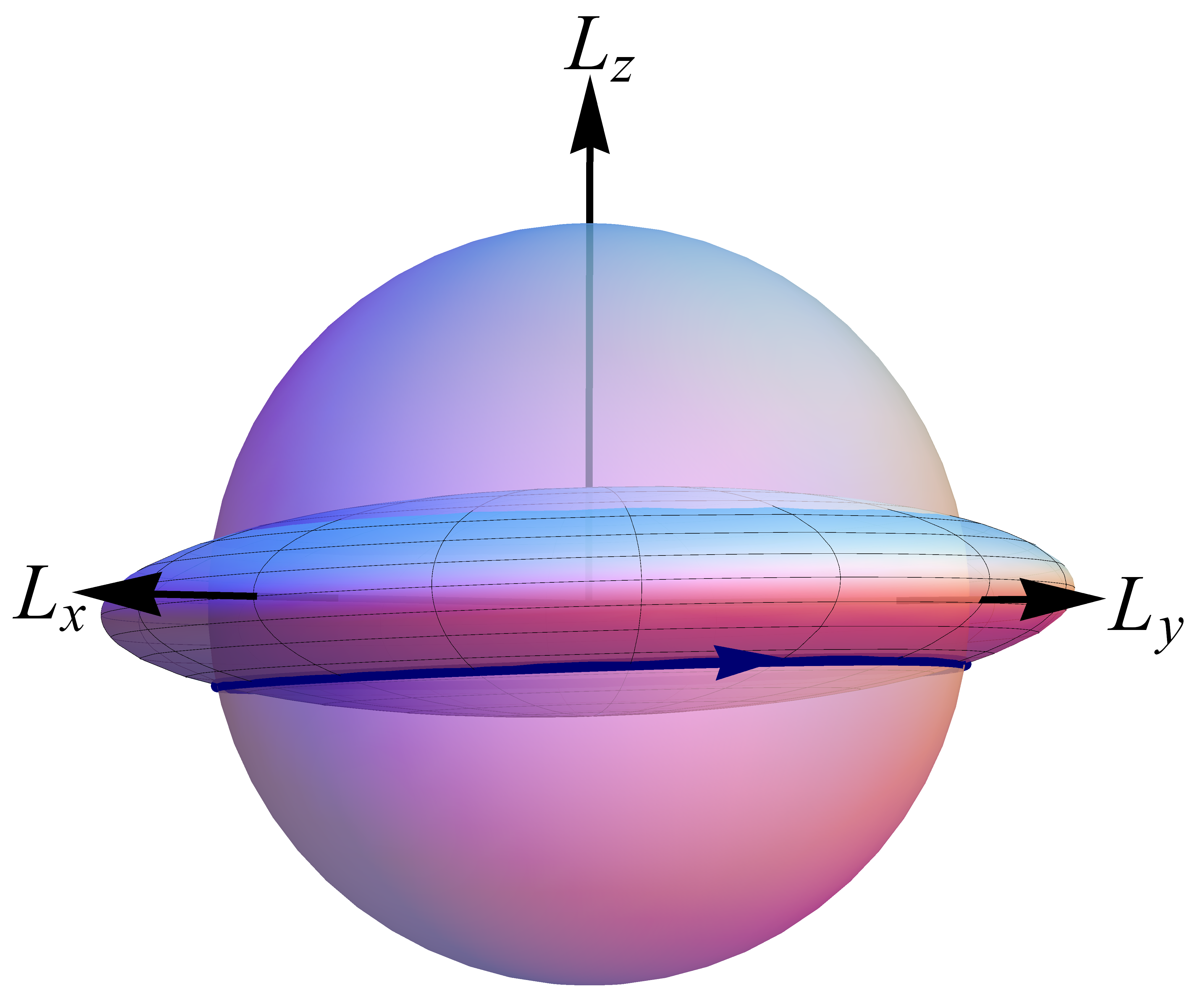}
\caption{The orbit $\bo{L}_3(t)$ corresponding to twisting somersaults in the counterclockwise direction.}
\label{fig:rigidorbit}
\end{figure}

Let $\bo{l}=(0,l,0)^t$ be the spatial angular momentum in $\mathcal{F}_S$, and suppose the athlete is rigid in twist position with shape DU so that the diagonalised tensor of inertia is $J_t = \diag(J_{t,x},J_{t,y},J_{t,z})$. Now if the initial angular momentum is 
\begin{equation}
\bo{L}_3(0)=\mathcal{R}_x(-\mathcal{X})\bo{l}\label{eq:L30}
\end{equation}
then once the quantity is rewritten in $\mathcal{F}_P$ the energy can be computed with \eqref{eq:conserveE} to obtain
\begin{equation}
E_3 = l^2\left(J_{t,y}^{-1}\cos^2{(p+\mathcal{X})}+J_{t,z}^{-1}\sin^2{(p+\mathcal{X})}\right)/2.
\end{equation}
With the tensor of inertia and energy known we can use \eqref{eq:Lorbit} to express the orbit as
\begin{equation}
\bo{L}_3(t)=R_p \bo{\mathcal{L}}(t; E_3, J_t, T_3/4),\label{eq:L3orbit}
\end{equation}
where $T_3=T(E_3,J_t)$ is the period of twist in the twisting somersault computed from \eqref{eq:T} and the phase shift constant $c=T_3/4$ is chosen to satisfy the initial condition \eqref{eq:L30}. In Figure \ref{fig:rigidorbit} we illustrate the orbit \eqref{eq:L3orbit} showing a slight tilt due to $R_p$, which is the rotation matrix needed to align $\mathcal{F}_P$ with $\mathcal{F}_C$.

For arbitrary energy $E$ and tensor of inertia $I$ the period $T(E,I)$ given by \eqref{eq:T} specifies the amount of time required to complete a loop around the $\bo{L}$-sphere. This corresponds to the rigid body physically completing a full twist with some amount of somersault $\Delta \phi$, which cannot be directly observed on the $\bo{L}$-sphere but can be recovered through reconstruction of the full dynamics. Montgomery \cite{Montgomery90} writes 
\begin{equation}
\Delta\phi = \frac{2ET}{l} - S \mod 2\pi \label{eq:mont}
\end{equation}
which partitions the change in orientation $\Delta\phi$ into the sum of the dynamic phase $2ET/l$ and geometric phase $S$. While Montgomery's formula is only applicable to rigid bodies, Cabrera \cite{Cabrera07} generalises the result to 
\begin{equation}
\Delta\phi =  
- S 
+\frac{1}{l}\int_0^T I^{-1}(\bo{L}-\bo{A})\cdot\bo{L}\,dt\mod 2\pi\label{eq:cab}
\end{equation}
for self-deforming bodies, where the tensor of inertia $I$ and momentum shift $\bo{A}$ are now functions of $t$, and $S$ is the solid angle 
enclosed by the orbit $\bo{L}$. As the momentum shift $\bo{A}$ is proportional to the shape velocities, Cabrera's formula \eqref{eq:cab} reduces to Montgomery's formula \eqref{eq:mont} in the absence of shape change. By keeping the surface area $A$ (not to be confused with the momentum shift $\bo{A}$) 
lying to the left of the oriented orbit $\bo{L}$ we have the relation to the solid angle $S=A/l^2$. To remove the $\operatorname{mod} 2\pi$ essential in distinguishing the amount of somersault, we appropriately define the surface area $A$ to be the area between $\bo{L}$ and the equator, see  \cite{TwistSom} and  
 \cite{WTongthesis} for more details. Intuitively, we expect more geometric phase in a faster twisting somersault because there is no geometric phase in the limiting case of no twist, and this is in agreement with our definition. The derivation of the solid angle $S$ can be found in \cite{TwistSom}, here we simply state
\begin{equation}
S(E/l^2,I) = 4 \sqrt{\frac{I_y}{I_x I_z(I_y-I_z)H_x}}\Big((I_x-I_z)\Pi(\nu,k^2)-I_x H_z K(k^2)\Big),\label{eq:S}
\end{equation}
where $k^2$ is given by \eqref{eq:tk} and
\begin{align}
H_x &= 2EI_x/l^2-1 & H_z&=1-2EI_z/l^2 & \nu &= -\frac{I_z(I_x-I_y)}{I_x(I_y-I_z)}.\nonumber
\end{align}

\section{Impulsive Shape Changes}\label{sec:fastkick}
\begin{figure}[b]
\centering
\subfloat[LU fast-kick\newline $\bo{L}_{{}_+} = \mathcal{R}_x(-\mathcal{X})\bo{L}_{{}_-}$]{\includegraphics[height=4.5cm]{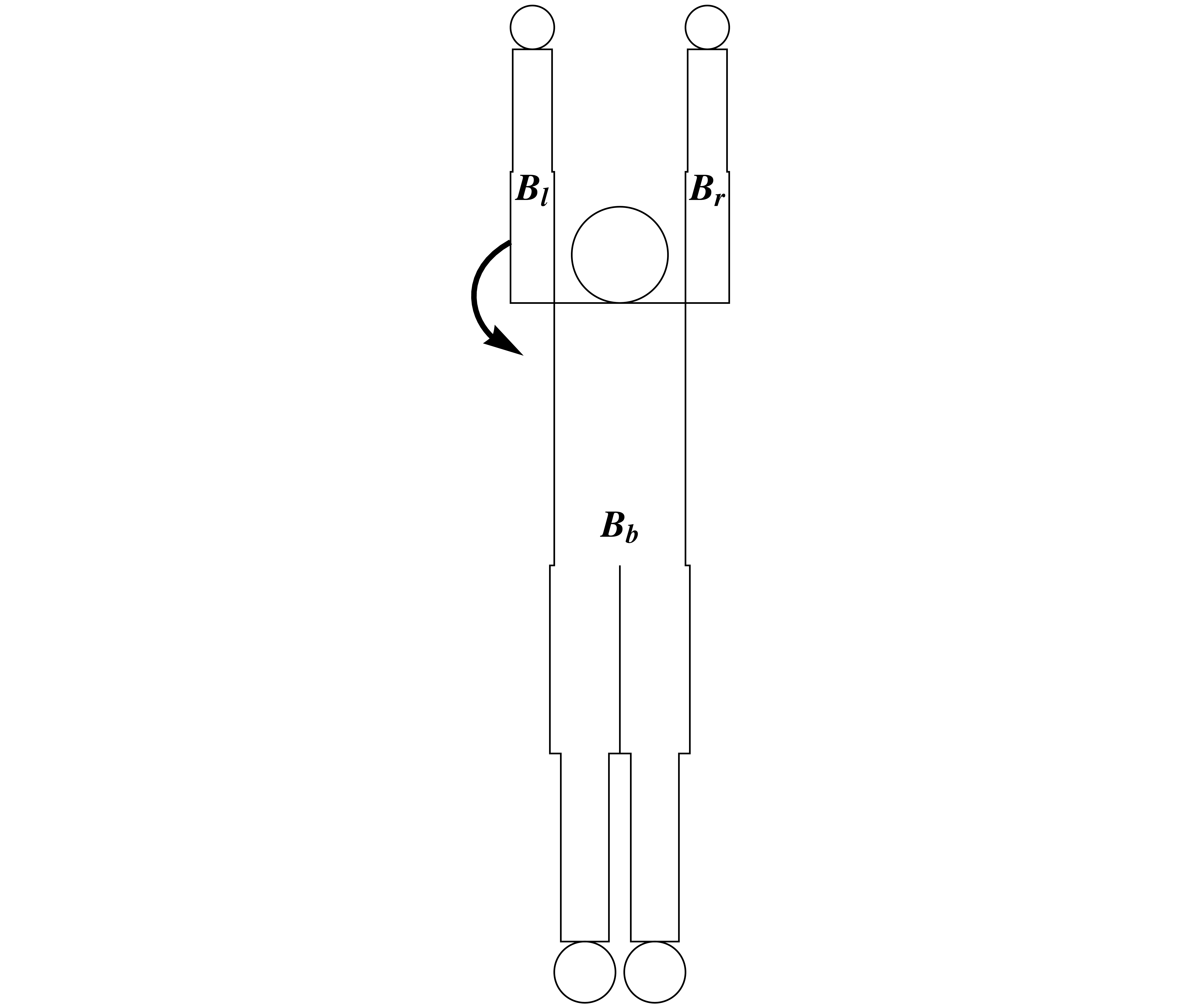}\label{newsubfig:kickLU}}
\subfloat[UL fast-kick\newline $\bo{L}_{{}_+} = \mathcal{R}_x(\mathcal{X})\bo{L}_{{}_-}$]{\includegraphics[height=4.5cm]{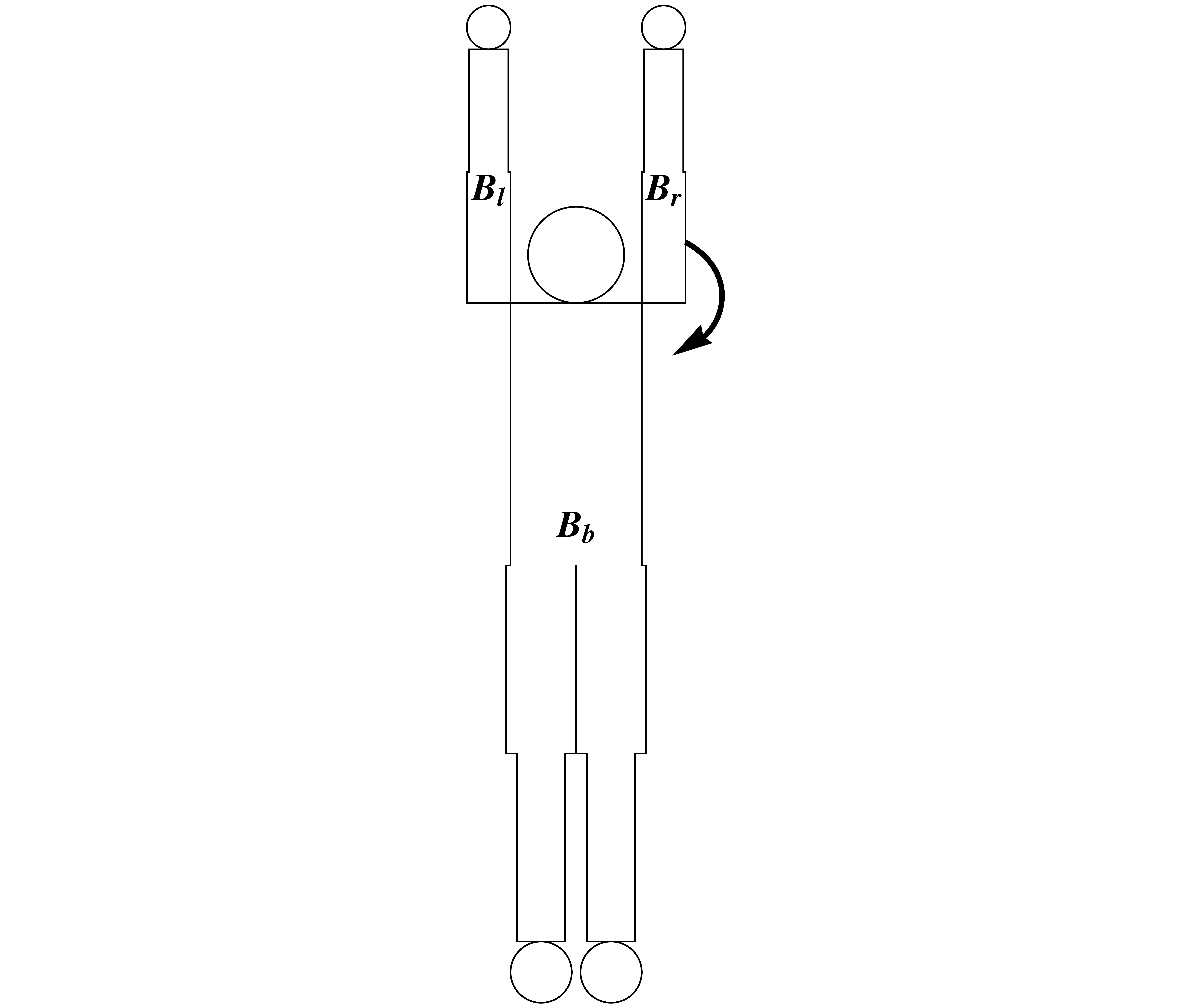}\label{newsubfig:kickUL}}
\subfloat[LH fast-kick\newline $\bo{L}_{{}_+} = \mathcal{R}_x(-\mathcal{Y})\bo{L}_{{}_-}$]{\includegraphics[height=4.5cm]{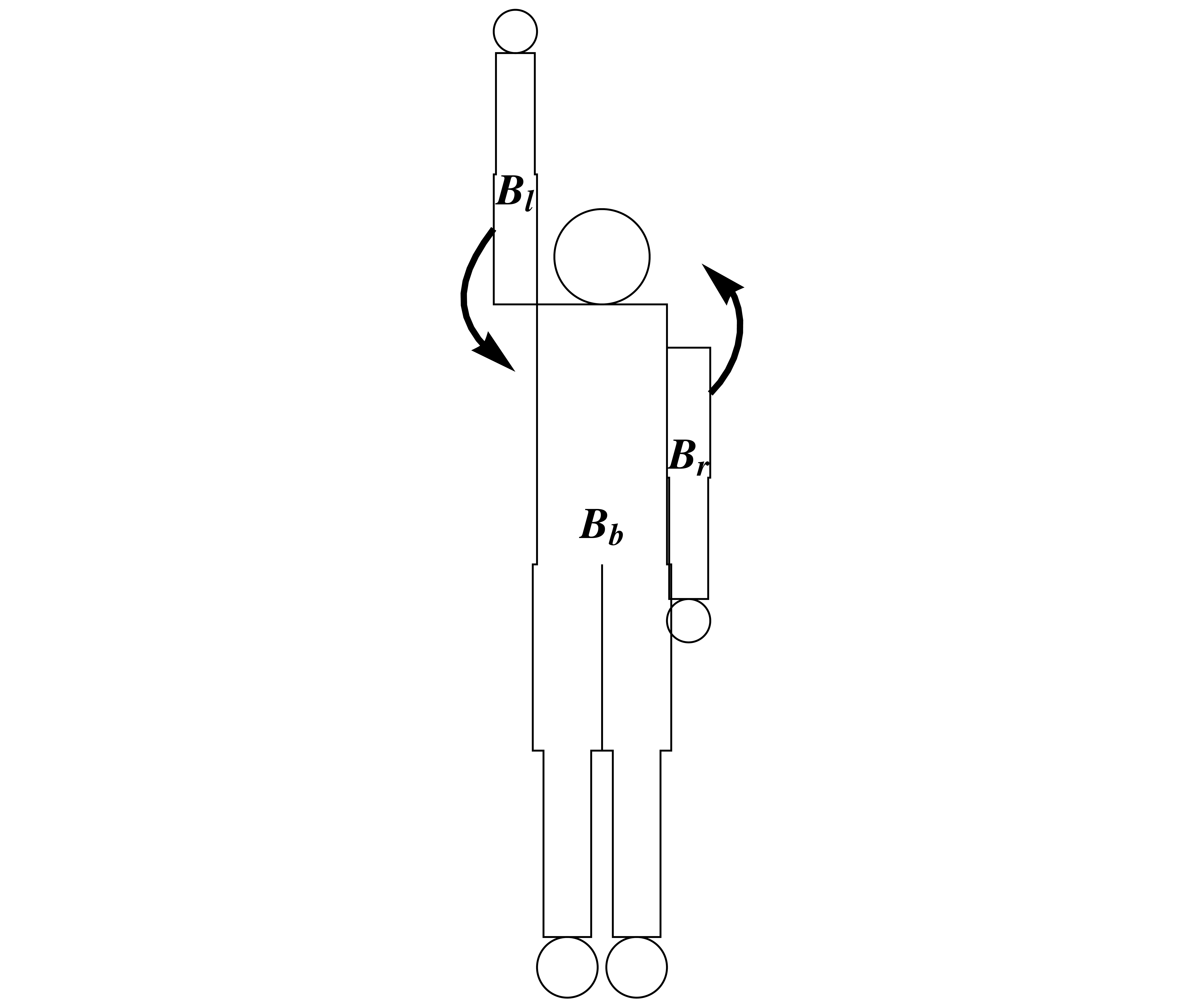}\label{newsubfig:kickLH}}\newline
\subfloat[HU fast-kick\newline $\bo{L}_{{}_+} = \mathcal{R}_x(\mathcal{X})\bo{L}_{{}_-}$]{\includegraphics[height=4.5cm]{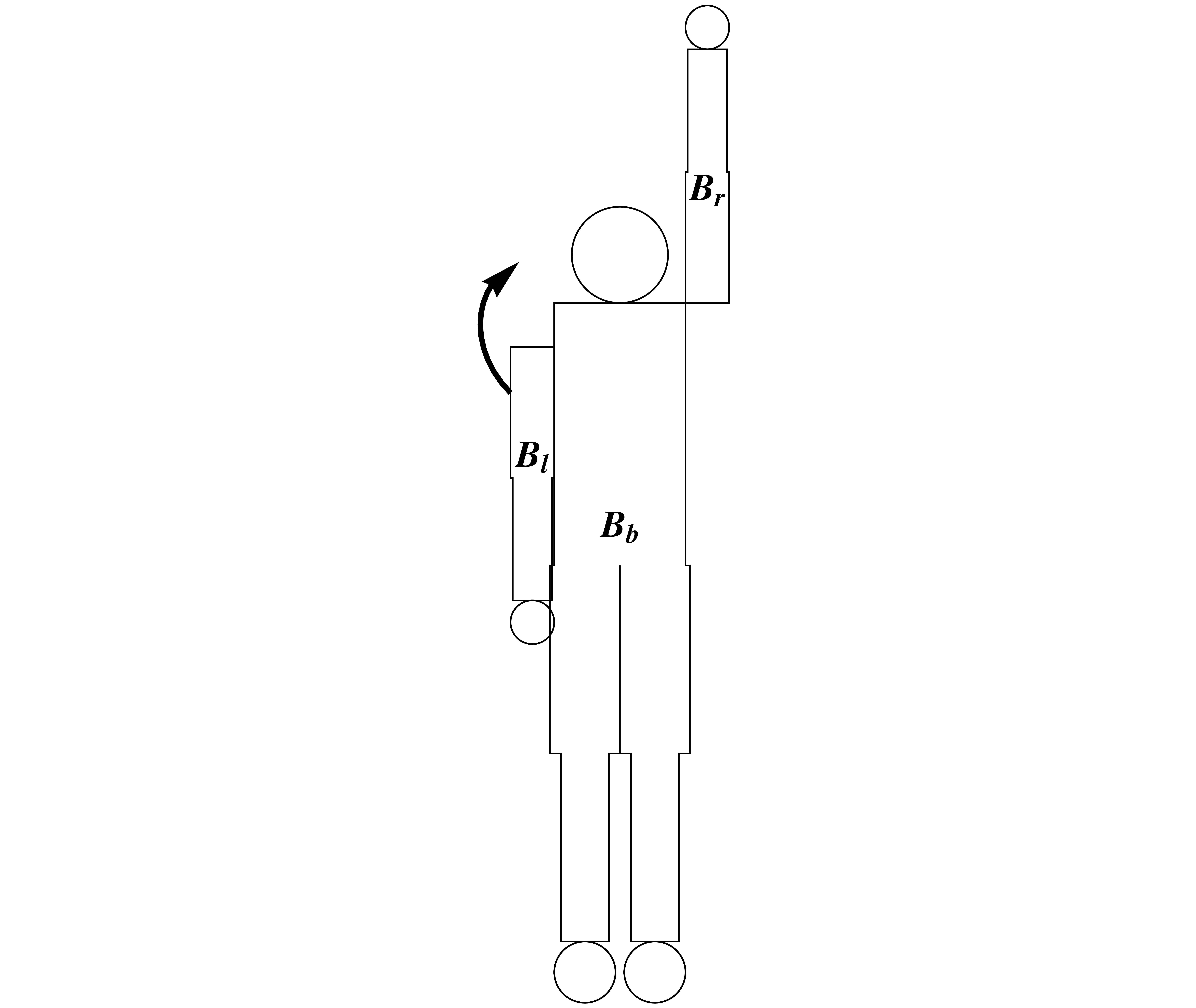}\label{newsubfig:kickHU}}
\subfloat[UH fast-kick\newline $\bo{L}_{{}_+} = \mathcal{R}_x(-\mathcal{X})\bo{L}_{{}_-}$]{\includegraphics[height=4.5cm]{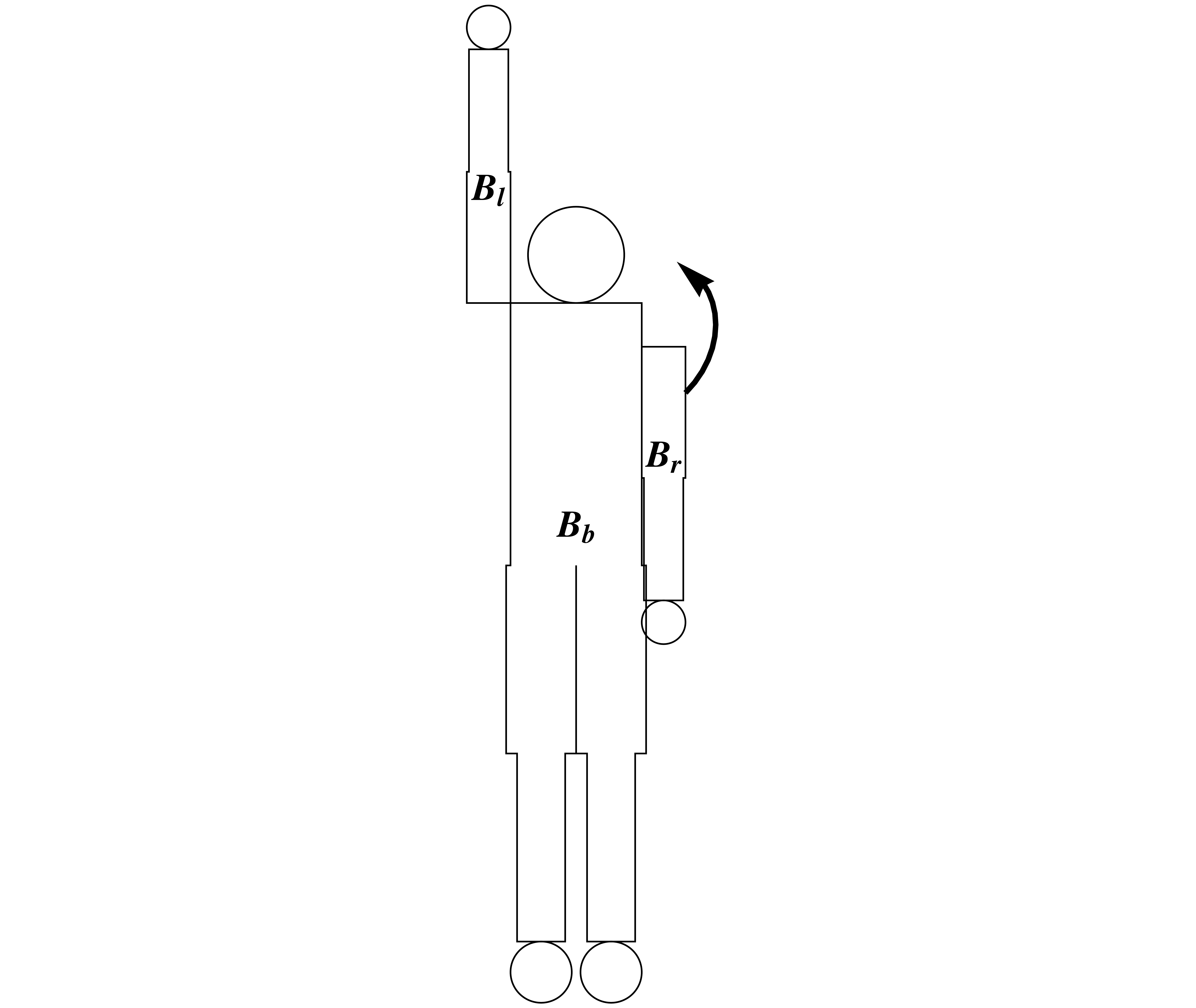}\label{newsubfig:kickUH}}
\subfloat[HL fast-kick\newline $\bo{L}_{{}_+} = \mathcal{R}_x(\mathcal{Y})\bo{L}_{{}_-}$]{\includegraphics[height=4.5cm]{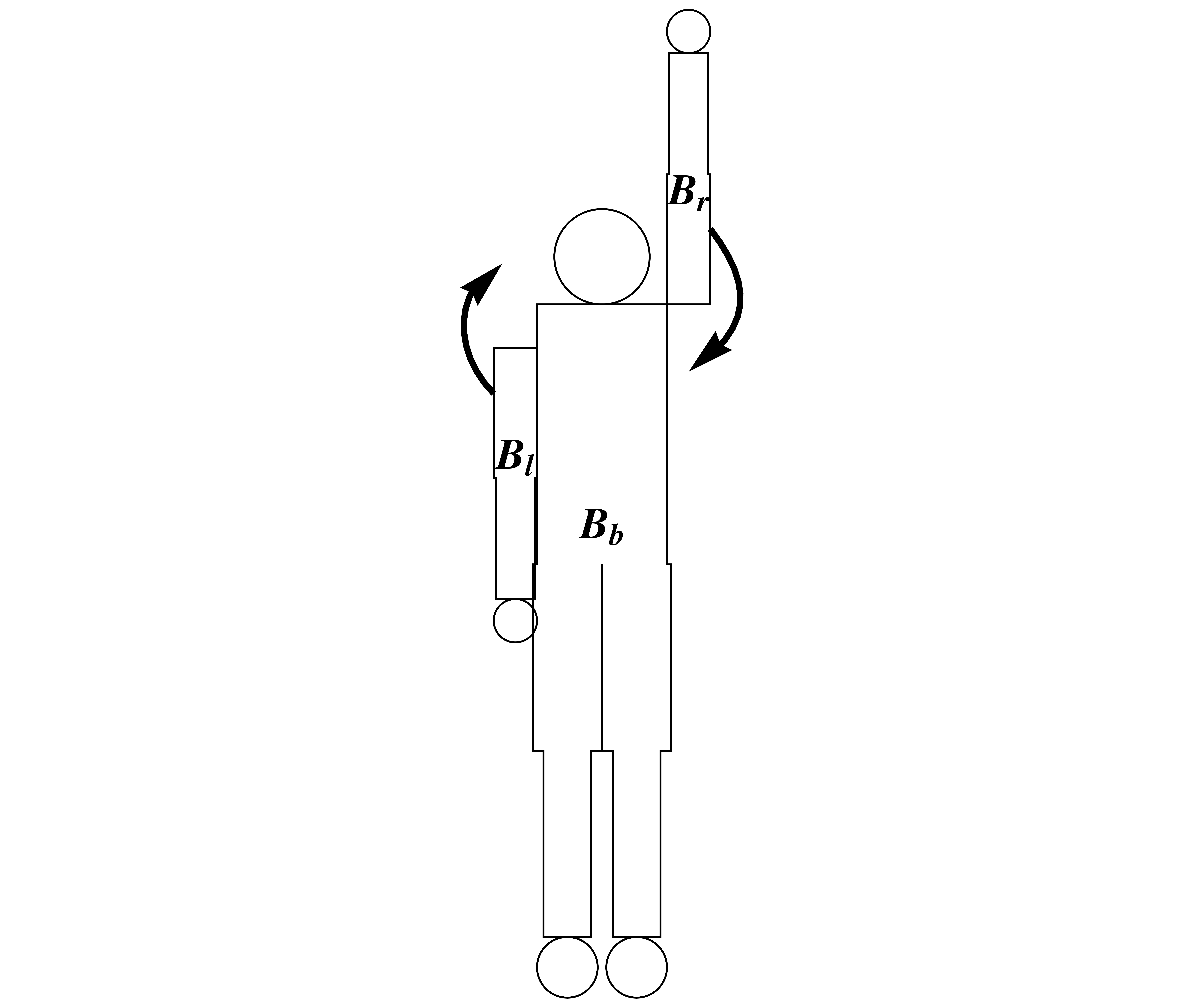}\label{newsubfig:kickHL}}
\caption{The tilt generated from different impulsive shape changes, where the computation details are specified in \cite{WTongthesis}.}\label{fig:fastkicks}
\end{figure}
When the shape change is instantaneous the equations of motion \eqref{eq:eom} simplify to 
\begin{equation}
\bo{\dot{L}}=I^{-1}\bo{A}\times\bo{L}.\label{eq:eomkick}
\end{equation}
This is because $\bo{L}$ remains at constant length while $\bo{A}$ diverges as it is proportional to the shape change velocities. When the shape change becomes impulsive we have $\bo{\Omega}=I^{-1}(\bo{L}-\bo{A})\rightarrow -I^{-1}\bo{A}$ resulting in \eqref{eq:eomkick}, and expressing the components using \eqref{eq:I3} and \eqref{eq:A3} gives
\begin{equation}
\bo{\dot{L}} = I_{xx}^{-1}(A_l\dot{\alpha}_l+A_r\dot{\alpha}_r)M\bo{L}
\end{equation}
where $M$ is a constant matrix. This is a time-dependent linear differential equation with solution
\begin{align}
\bo{L}_{{}_+} &= \exp{(\Theta M)}\bo{L}_{{}_-} = \mathcal{R}_x(\Theta)\bo{L}_{{}_-},\label{eq:Lkick}
\end{align}
where $\bo{L}_{{}_-}$ and $\bo{L}_{{}_+}$ are the instantaneous angular momenta before and after the shape change, and the angle
\begin{figure}[t]
\centering\begin{tabular}{>{\centering\arraybackslash}m{1.5cm} >{\centering\arraybackslash}m{1cm} >{\centering\arraybackslash}m{1.5cm} >{\centering\arraybackslash}m{1cm} >{\centering\arraybackslash}m{1.5cm} >{\centering\arraybackslash}m{1cm} >{\centering\arraybackslash}m{1.5cm} >{\centering\arraybackslash}m{1cm} >{\centering\arraybackslash}m{1.5cm}} 
UU&   &DU&   &UU\\
\includegraphics[width=1.5cm]{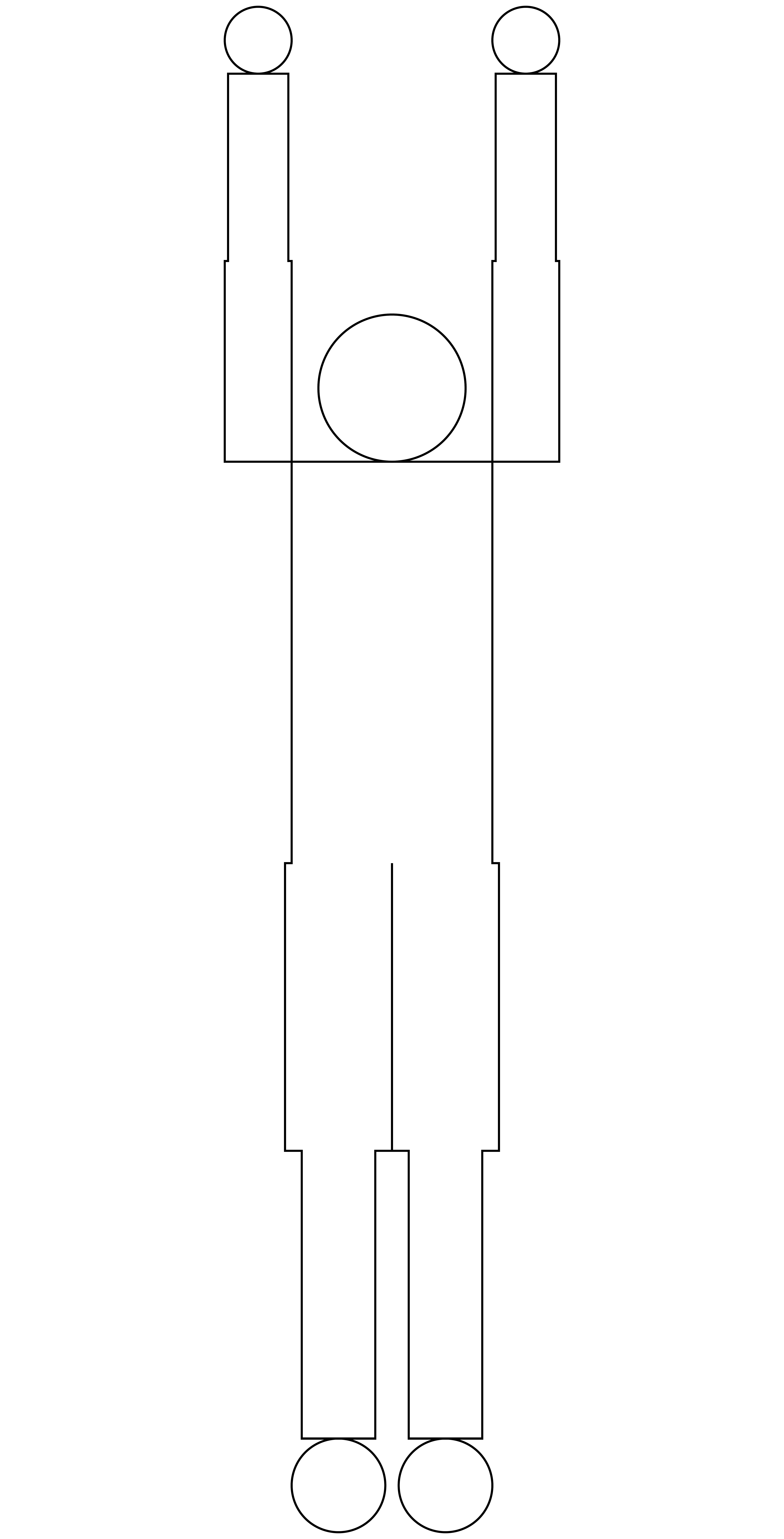} & $\begin{array}{c}\text{LU}\\ \longrightarrow \end{array}$ & \includegraphics[width=1.5cm]{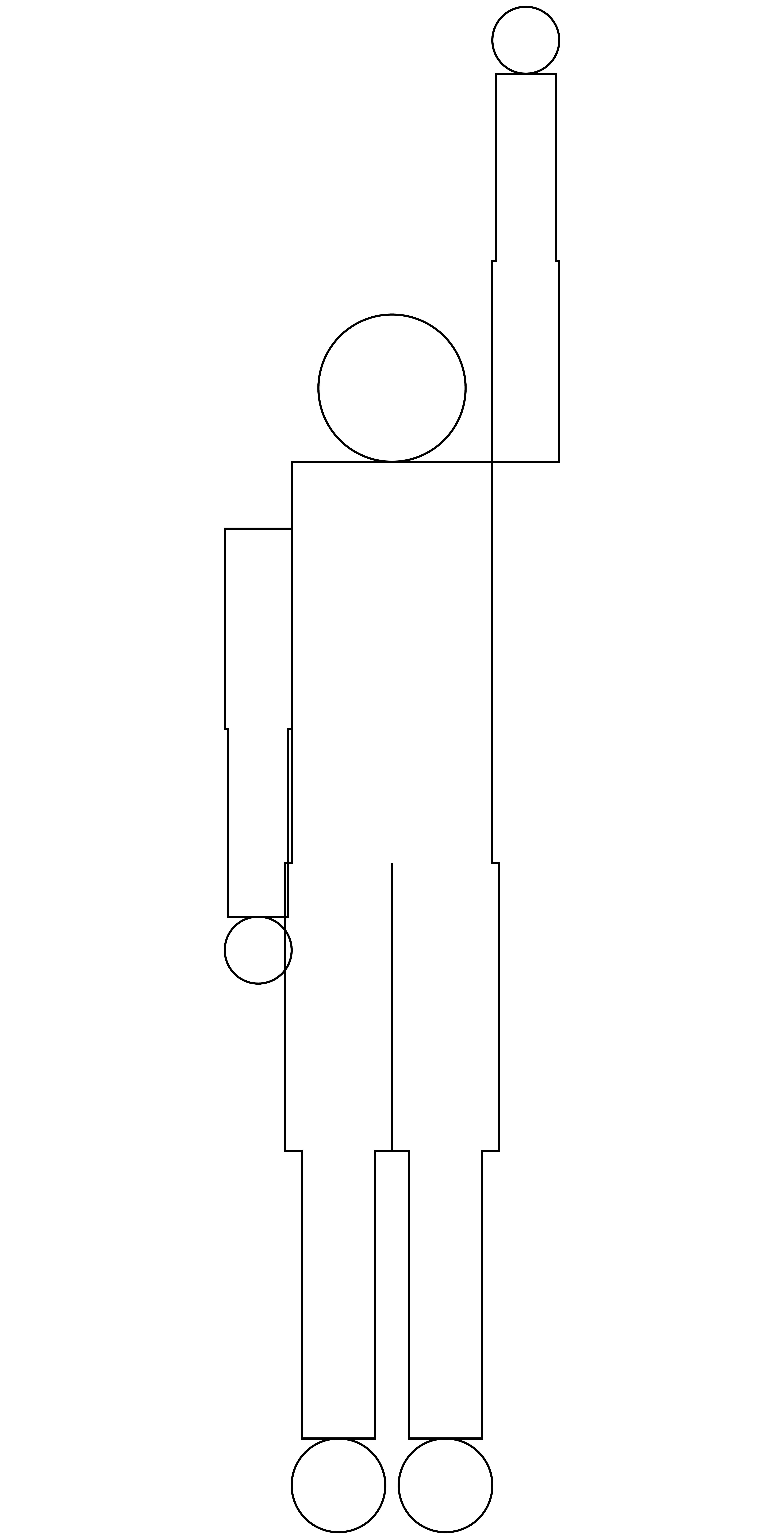} & $\begin{array}{c}\text{HU}\\ \longrightarrow\end{array}$ & \includegraphics[width=1.5cm]{UU.pdf}
\end{tabular}
\caption{Simplest 5-stage dive mechanic for executing twisting somersaults.}
\label{fig:5stage}
\end{figure}
\begin{table}[b]\centering
\begin{tabular}{cccc}
\hline
Stage & Energy $E_i$ & Somersault rate 	& Time spent $\tau_i$\\
Shape & Period $T_i$ & Twist rate				& Orbit $\bo{L}_i(t)$\\\hline\hline
1 		& $0.0243l^2$ 	& $0.04852 l$			& $(64.7452 m-19.4154 n)/l$\\
UU		&	$\infty$			& $0$							& $(0,l,0)^t$\\\hline\hline
3			& $0.0452l^2$		& $0.05548 l$			& $33.9610n/l$\\
DU		& $33.9610/l$		& $0.18501 l$			& $R_p \bo{\mathcal{L}}(t; E_3, J_t, T_3/4)$\\\hline\hline
5 		& $0.0243l^2$ 	& $0.04852 l$			& $(64.7452 m-19.4154 n)/l$\\
UU		&	$\infty$			& $0$							& $(0,l,0)^t$\\\hline\hline
&&\multicolumn{2}{c}{Total time $\mathcal{T}_5=(129.4905m - 4.8699n)/l$}\\\cline{3-4}
\end{tabular}
\caption{The simple twisting somersault consisting of $m$ somersaults and $n$ twists that follows the 5-stage dive mechanic given by Figure \ref{fig:5stage}.}
\label{tab:5stage}
\end{table}
\begin{figure}[t]
\subfloat[$\bo{L}$-sphere.]{\includegraphics[width=6.9cm]{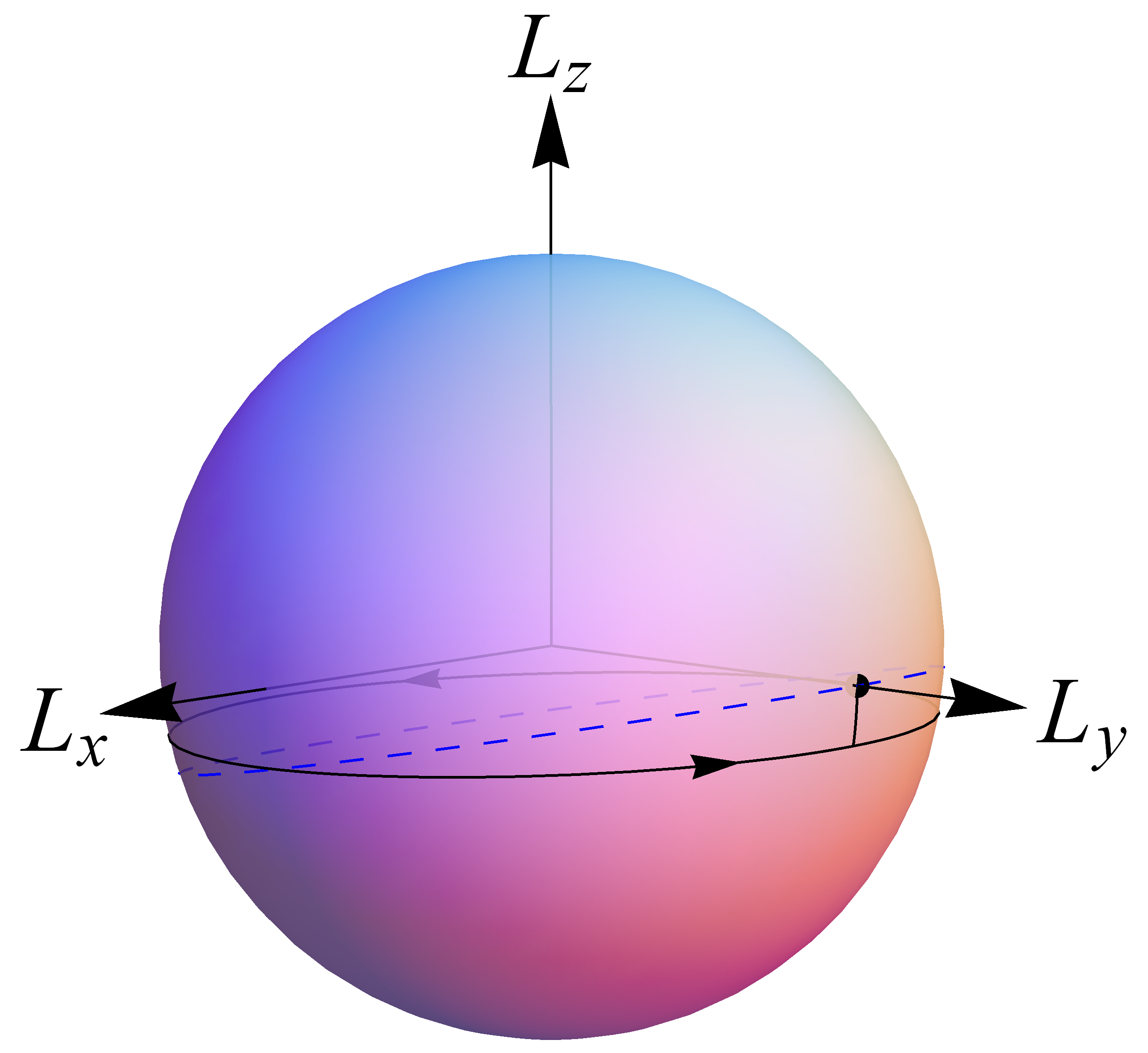}}
\subfloat[Mercator projection.]{\includegraphics[width=6.9cm]{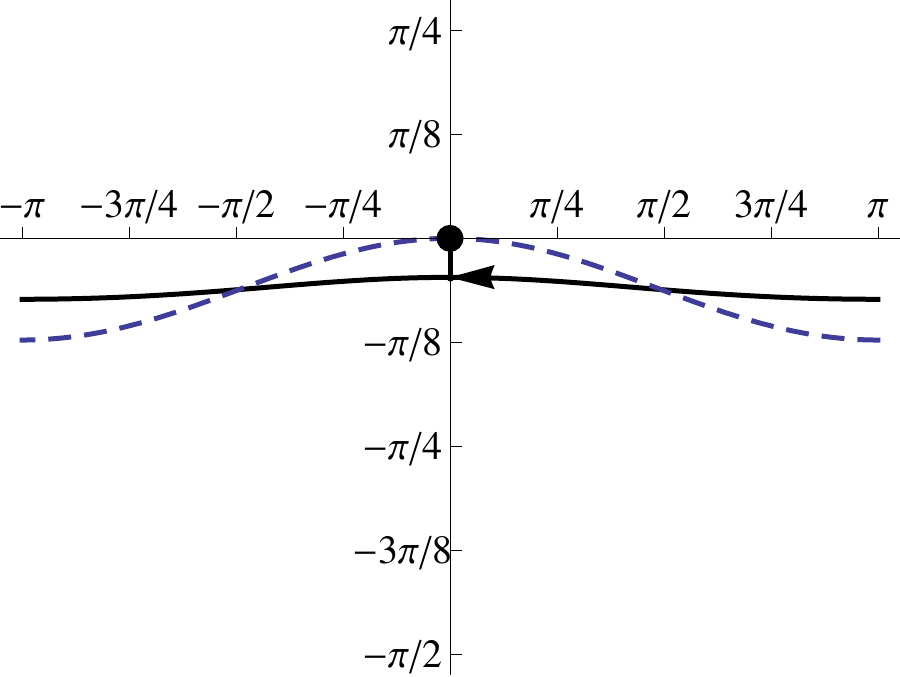}}
\caption{The point shown on the $L_y$-axis is the steady rotation corresponding to pure somersaulting motion. The black solid loop shows the orbit $\bo{L}_3(t)$ and the blue dashed loop represents the family of initial conditions $\bo{L}_5(0;\tau_3)$.}
\label{fig:L3kick}
\end{figure}
\begin{figure}[b]
\includegraphics[width=8cm]{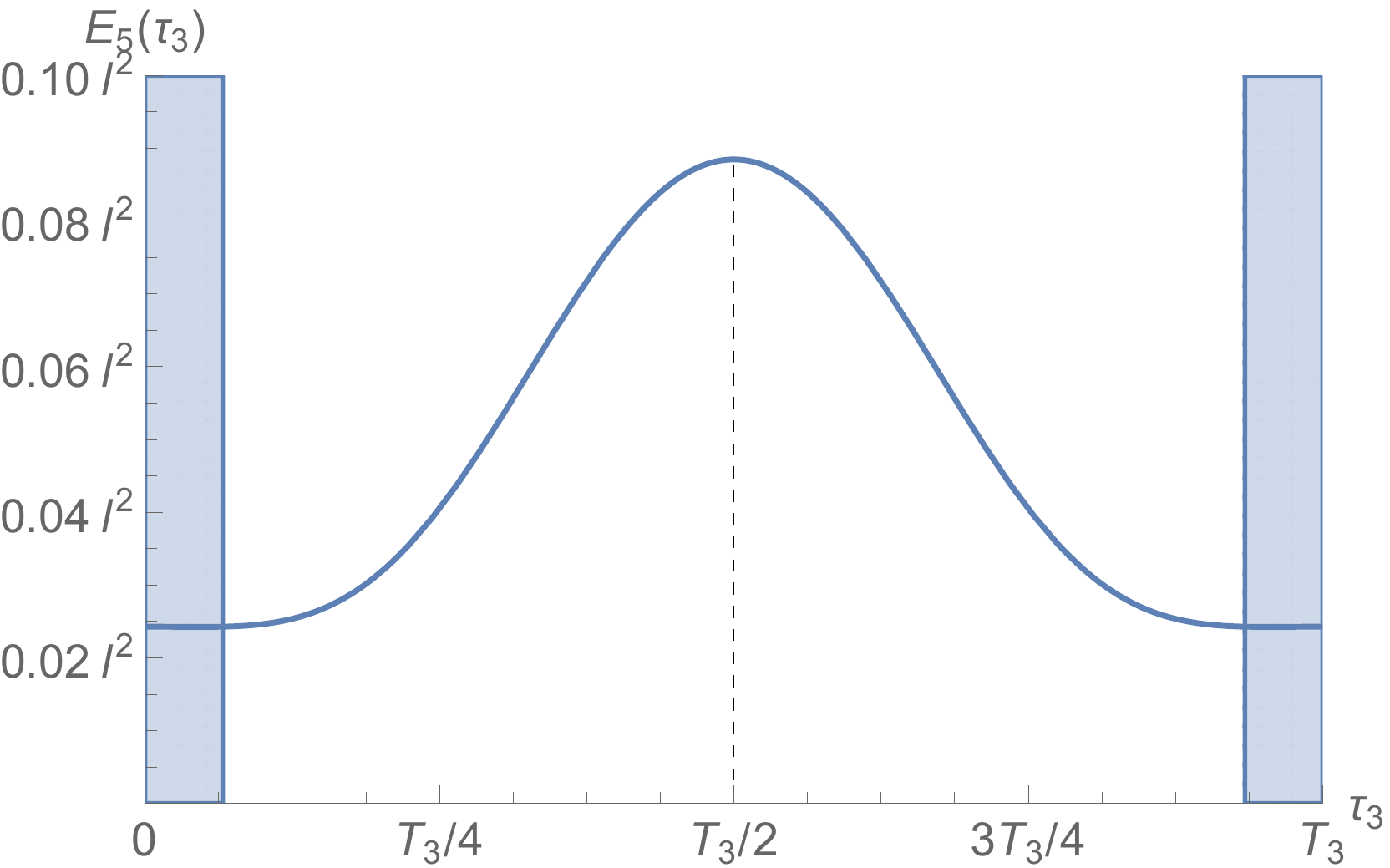}
\caption{The energy of stage 5 that results from the impulsive HU shape change expressed as a function of time spent in stage 3. The blue region denotes $E_5 < l^2/(2 I_{s,y})$, which results in wobbling somersaulting motion. The period of wobble becomes infinite when $\tau_3=0\mod T_3$, and this yields the familiar pure somersaulting motion of the athlete.}
\label{fig:E5tk}
\end{figure}
\begin{figure}[t]
\centering
\begin{tabular}{>{\centering\arraybackslash}m{1.5cm} >{\centering\arraybackslash}m{1cm} >{\centering\arraybackslash}m{1.5cm} >{\centering\arraybackslash}m{1cm} >{\centering\arraybackslash}m{1.5cm} >{\centering\arraybackslash}m{1cm} >{\centering\arraybackslash}m{1.5cm} >{\centering\arraybackslash}m{1cm} >{\centering\arraybackslash}m{1.5cm}} 
UU&   &DU&   &UD&   &DU&   &UU\\
\includegraphics[width=1.5cm]{UU.pdf} & $\begin{array}{c}\text{LU}\\ \longrightarrow \end{array}$ & \includegraphics[width=1.5cm]{DU.pdf} & $\begin{array}{c}\text{HL}\\ \longrightarrow\end{array}$ & \includegraphics[width=1.5cm]{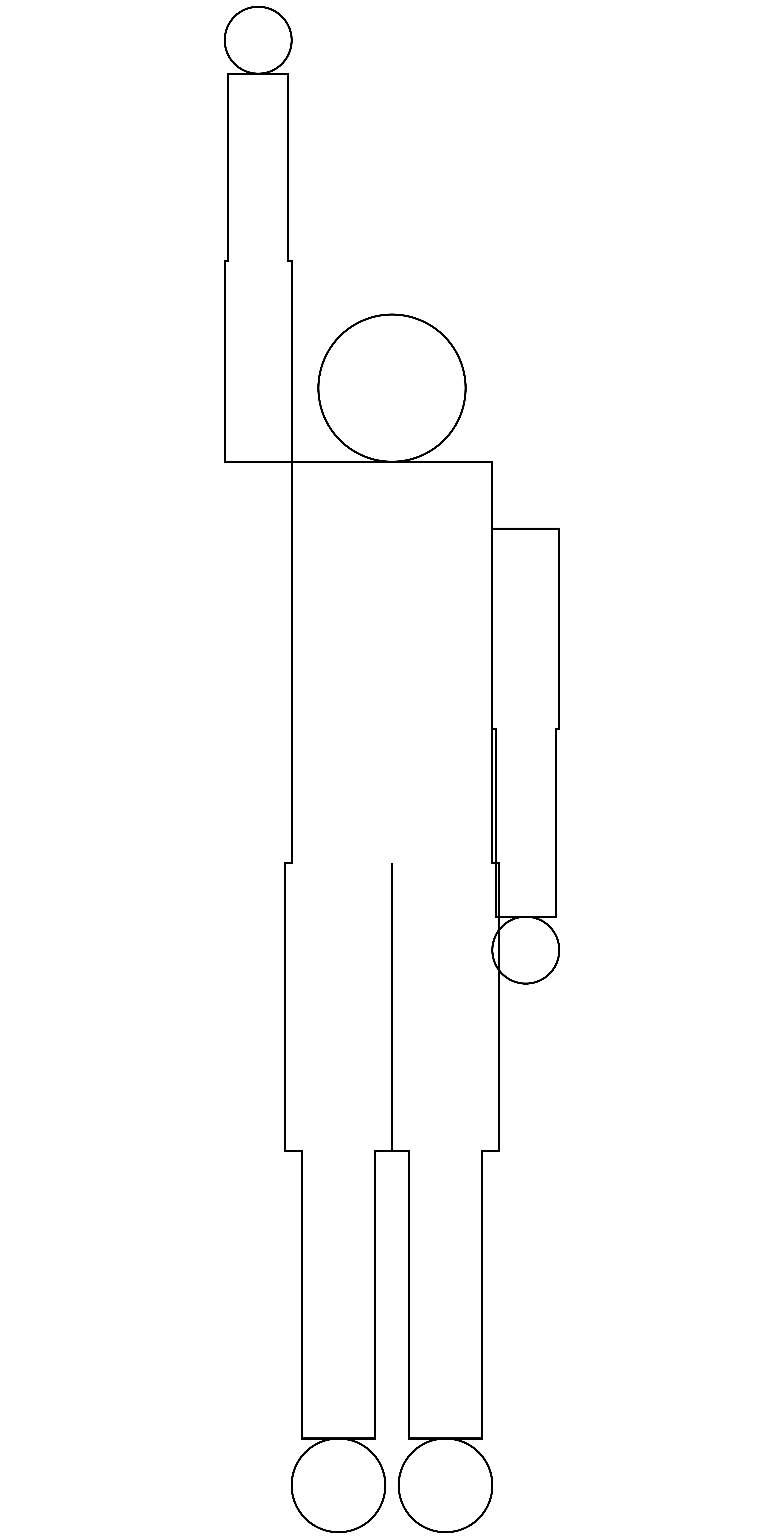}& $\begin{array}{c}\text{LH}\\ \longrightarrow\end{array}$ & \includegraphics[width=1.5cm]{DU.pdf}& $\begin{array}{c}\text{HU}\\ \longrightarrow\end{array}$ & \includegraphics[width=1.5cm]{UU.pdf}
\end{tabular}
\caption{9-stage dive that enables athletes to execute more twists.}
\label{fig:9stage}
\end{figure}
\begin{figure}[b]
\subfloat[$\bo{L}$-sphere.]{\includegraphics[width=6.9cm]{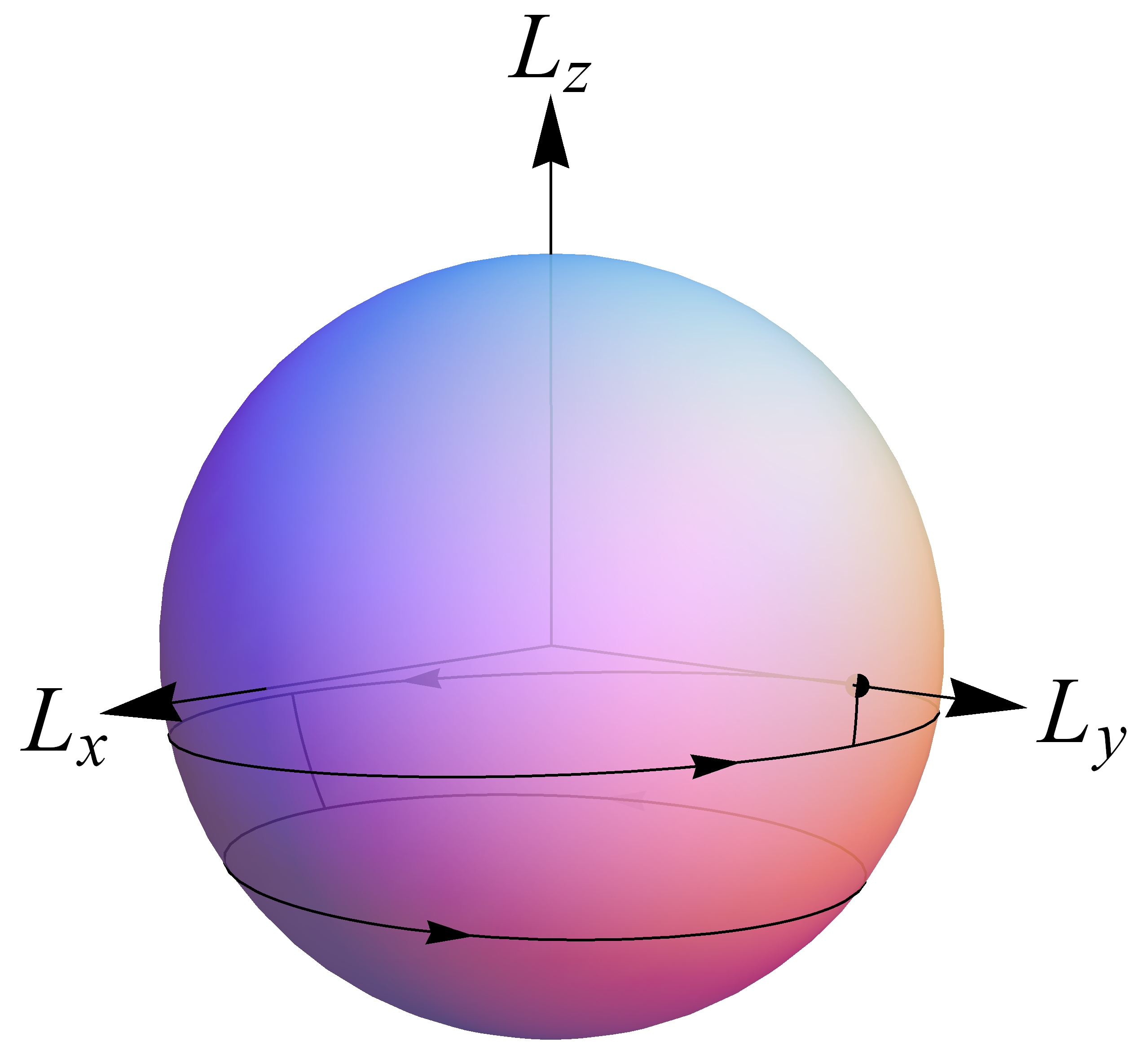}}
\subfloat[Mercator projection.]{\includegraphics[width=6.9cm]{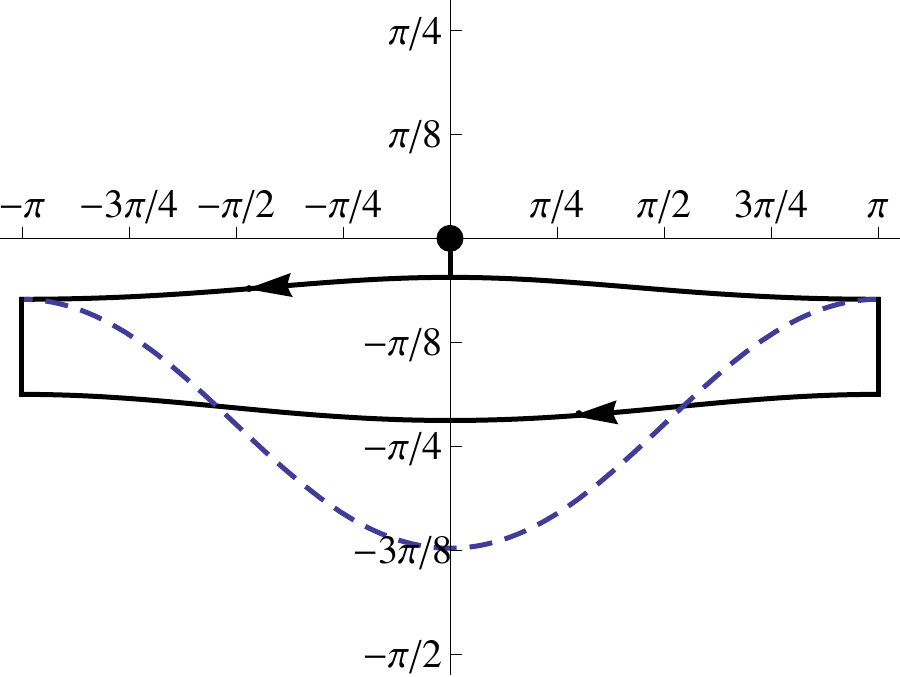}}
\caption{The LU fast-kick takes the orbit from the equilibrium point to the upper loop, and the HL fast-kick brings the orbit to the lower loop corresponding to a faster twisting somersaulting state. The LH and HU fast-kicks are then used to reverse this procedure so that the orbit returns to the equilibrium point, which corresponds to the athlete completing the dive in pure somersaulting motion.}
\label{fig:9stagekick}
\end{figure}

\begin{equation}
\Theta=\displaystyle\lim_{\tau\rightarrow 0^+}\int_0^\tau I_{xx}^{-1}(A_l \dot{\alpha}_l+A_r\dot{\alpha}_r)\,dt\label{eq:tilt}
\end{equation}
is the tilt generated from the impulsive shape change. We appropriately parameterise the shape angles $\alpha_l$ and $\alpha_r$ to evaluate \eqref{eq:tilt}, and for the impulsive shape changes shown in Figure \ref{fig:fastkicks} the numerical values are $\mathcal{X}\approx 0.147$ and $\mathcal{Y}\approx 0.330$.

Combining rigid body orbit \eqref{eq:Lorbit} with \eqref{eq:Lkick} for impulsive shape changes we can now formulate twisting somersaults. Following the construction of \cite{TwistSom} the simplest twisting somersault dive mechanic is shown in Figure \ref{fig:5stage}, where the athlete takes off and finishes the dive in pure somersaulting motion while using LU to initiate and HU fast-kick to terminate twisting motion mid-flight. In the initial and final stages of the dive the orbit is
\begin{equation}
\bo{L}_1(t)=\bo{L}_5(t)=\bo{l}=(0,l,0)^t,
\end{equation}
which corresponds to an equilibrium point on the $\bo{L}$-sphere, while in stage 3 the orbit is given by \eqref{eq:L3orbit} and is illustrated in Figure \ref{fig:rigidorbit}. The technical details of the dive consisting of $m$ somersaults and $n$ twists are given in \cite{TwistSom}, with the dive summary provided in Table \ref{tab:5stage}. In the table we set $\tau_1=\tau_5$ and use the $m$ somersault constraint in \eqref{eq:mont} to determine the total time $\tau_1+\tau_5$ spent in pure somersaulting motion. 
The entries in Table~\ref{tab:5stage} called ``Somersault rate'' and ``Twist rate'' give the amount the somersault and twist angle increases per second, respectively.

Now we are going to develop ideas that will lead towards the new 513XD dive.
An interesting point to note is that when the HU fast-kick in stage 4 is performed at a different time it has a different effect. 
Performing the kick at time $\tau_3$ leads to a point we denote by $\bo{L}_5(0;\tau_3)$, which will then be taken as the initial point of stage 5 in the 9-stage dive.
The curve  $\bo{L}_5(0;\tau_3)$ is shown as a dashed curve in Figure \ref{fig:L3kick}. From this figure it is already clear that if we perform the kick 
near the point opposite to the initial point $\bo{L}_1$ it has the opposite effect: instead of decreasing $L_z$ it increases $L_z$.
A larger $L_z$ implies a faster twisting motion, and hence the idea is born to try to make this rotation as fast as possible.
To determine the fastest possible twisting motion we plot
the energy $E_5(\tau_3)$ of stage 5 expressed as a function of time $\tau_3$ when the kick is performed in stage 3, see 
 Figure \ref{fig:E5tk}.
 The maximum possible energy is $E_5(T_3/2) = 0.0885l^2$. This minimises the period of twist and allows the athlete to perform additional twists. 
 To further improve the faster twisting somersault rate we can replace the HU fast-kick with HL, which uses shape change from both arms to essentially double the effect. Figure \ref{fig:9stage} presents a 9-stage dive sequence that takes advantage of this result, and the orbit of the faster twisting somersault is shown in Figure \ref{fig:9stagekick}. In order to use \eqref{eq:mont} to compute the total somersault amount given $n$ twists, we combine the nine stages to obtain
\begin{equation}
\Delta\phi_\text{kick}=\sum_{{i=1}\atop {i \text{ odd}}}^9\frac{2E_i\tau_i}{l}-S(E_3/l^2,J_t)-(n-1)S(E_5/l^2,J_t),
\end{equation}
where the solid angle $S$ is computed using \eqref{eq:S}, energy $E_i$ with \eqref{eq:conserveE} and $\tau_i$ by \eqref{eq:T} for the twisting somersaulting motions. We can then set $\Delta\phi_\text{kick}=2\pi m$ for $m$ somersaults and solve for the total pure somersaulting time $\tau_1+\tau_9$, and provide the 9-stage dive summary in Table \ref{tab:9stage}.
\begin{table}[b]\centering
\begin{tabular}{cccc}
\hline
Stage & Energy $E_i$  & Somersault rate & Time spent $\tau_i$\\
Shape & Period $T_i$  & Twist rate			& Orbit $\bo{L}_i(t)$\\\hline\hline
1 		& $0.0243l^2$   & $0.04852 l$			& $(64.7452 m - 6.5082 n-12.9073)/l$\\
UU		&	$\infty$			& $0$							& $(0,l,0)^t$\\\hline\hline
3			& $0.0452l^2$		& $0.05548 l$			& $16.9805/l$\\
DU		& $33.9610/l$		& $0.18501 l$			& $R_p \bo{\mathcal{L}}(t; E_3, J_t, T_3/4)$\\\hline\hline
5 		& $0.1835l^2$ 	& $0.05547 l$			& $11.3854(n-1)/l$\\
UD		&	$11.3854/l$		& $0.55187 l$			& $R_p^{-1} \bo{\mathcal{L}}(t; E_5, J_t, 3T_5/4)$\\\hline\hline
7			& $0.0452l^2$		& $0.05548 l$			& $16.9805/l$\\
DU		& $33.9610/l$		& $0.18501 l$			& $R_p \bo{\mathcal{L}}(t; E_3, J_t, 3T_3/4)$\\\hline\hline
9 		& $0.0243l^2$   & $0.04852 l$			& $(64.7452 m - 6.5082 n-12.9073)/l$\\
UU		&	$\infty$			& $0$							& $(0,l,0)^t$\\\hline\hline
&&\multicolumn{2}{c}{Total time $\mathcal{T}_9=(129.4905 m - 1.6310 n-3.2389)/l$}\\\cline{3-4}
\end{tabular}
\caption{The innovative twisting somersault consisting of $m$ somersaults and $n$ twists that follows the 9-stage dive mechanic given by Figure \ref{fig:9stage}.}
\label{tab:9stage}
\end{table}

Comparison of Table \ref{tab:5stage} with Table \ref{tab:9stage} shows that for $n>1$ twists the diver completes the twists significantly faster with the 9-stage dive mechanic, resulting in more time spent in pure somersaulting motion to complete the desired $m$ somersaults. For the total times of the 5-stage and the 9-stage dive we have $\mathcal{T}_5 < \mathcal{T}_9$ when $n>1$.
For both dives it appears counterintuitive that performing more twists will take less time.
This is related to the fact that our somersault is performed in layout position and hence $J_{t,y}<I_{s,y}$.
Using a more complicated model allowing the athlete to enter pike or tuck position for the pure somersault would reverse the inequality between the moments of inertia,
and hence lead to dives which take longer when increasing $n$.
We also observe in Table \ref{tab:9stage} that the somersaulting rate of stage 5 is only slightly slower than stage 3 (and stage 7), while the twisting rate nearly triples as a result of $J_{t,x}\approx J_{t,y} \gg J_{t,z}$. 

For given $m$ and $n$ the time for each stage as listed in Tables~\ref{tab:5stage} and~\ref{tab:9stage} must be positive, otherwise the dive with the given $m$ and $n$ is impossible, even 
in the kick model. For the 5-stage dive (Table~\ref{tab:5stage}) the maximal $n$ for $m=1.5$ is 5, while for the 9-stage dive (Table~\ref{tab:9stage}) the maximal $n$ for $m=1.5$ is 12.
These numbers for $n$ are unrealistically high, which is a result of the kick-approximation. In reality the shape change takes a considerable amount of time.
In the next section we use our theory together with numerical integration to show that 5 twists 
can be performed with realistic shape changing times in a 9-stage dive.


\section{The 513XD Dive}\label{sec:513XD}
We now present the 513XD dive consisting of 1.5 somersaults and 5 twists with realistic shape changes building on the mechanism described in the previously section. The dive mechanics are given in Figure \ref{fig:9stage} as before, but we will now resort to numerics when solving the equations of motion \eqref{eq:eom} during shape change. For the 513XD dive simulation we choose $l=100$, $\tau_{i\text{ even}}=1/4$ and use a cubic spline for the shape angles to ensure the velocities are continuous throughout the dive. A video of the simulation of the 513XD dive along with some others is available at the first author's web page \url{http://www.maths.usyd.edu.au/u/williamt}.
\begin{figure}[b]
\subfloat[Energy $E_5$ as a function of $\tau_3$, where the maximum is $E_5=1801.23$ occurring at $\tau_3=0.0564T_3$.]{\includegraphics[width=7.4cm]{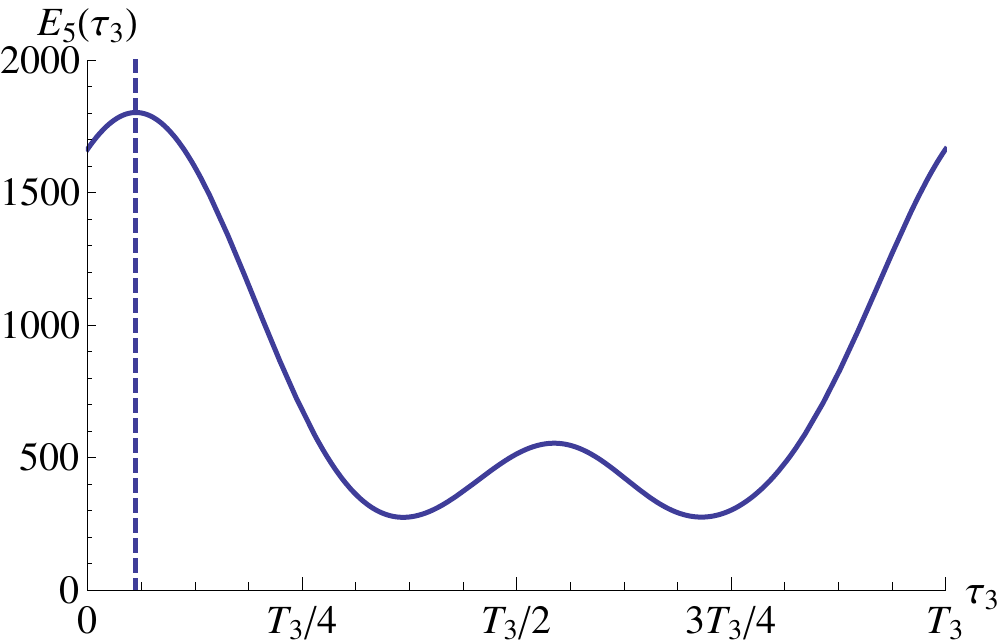}\label{fig:E5slow}}
\subfloat[Energy $E_7$ as a function of $\tau_5$, where the minimum is $E_7=460.012$ occurring at $\tau_5=0.7984T_5$.]{\includegraphics[width=7.4cm]{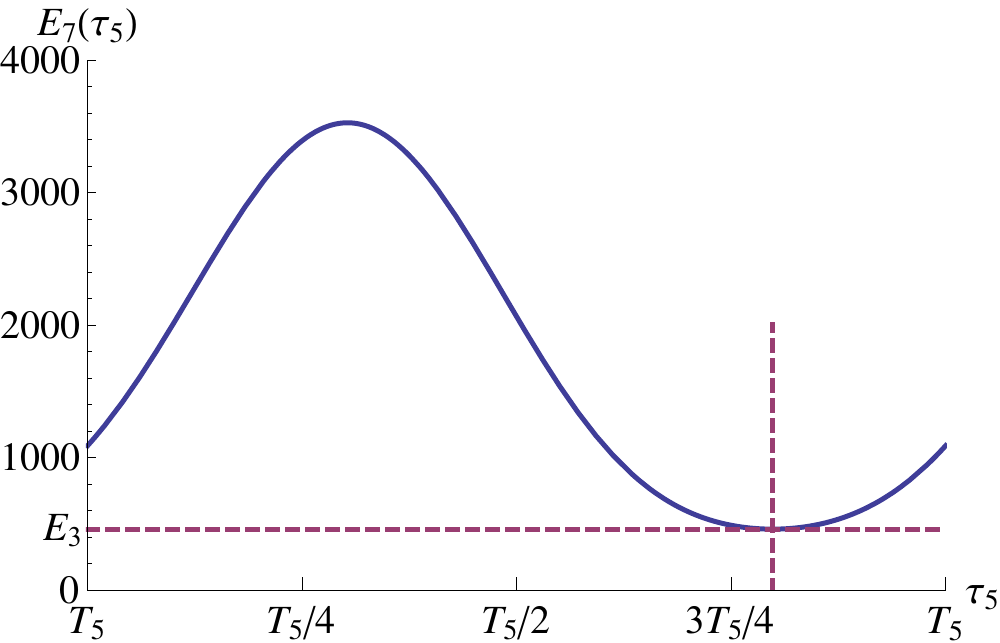}\label{fig:E7slow}}
\caption{The family of possible energies moving into and out of the faster twisting somersaulting motion. The vertical dashed lines show the timing of maximal and minimal energies, respectively.}\label{fig:slowenergies}
\end{figure}

The diver takes off with $\bo{L}_1(t)$ as before, and the trajectory $\bo{L}_2(t)$ is obtained by solving \eqref{eq:eom} numerically. We establish the rigid body orbit $\bo{L}_3(t)$ with \eqref{eq:Lorbit}, which differs from \eqref{eq:L3orbit} because $\bo{L}_2(\tau_2)$ determines the energy and phase shift instead of $\mathcal{R}_x(\mathcal{-X})\bo{L}_1(\tau_1)$ as previously. 
The trajectory $\bo{L}_4(t)$ is obtained by solving \eqref{eq:eom} numerically with initial condition $\bo{L}_3(\tau_3)$, meaning the end point $\bo{L}_4(\tau_4)$ can be used to compute the next stage's energy with different timing $\tau_3$ (see Figure \ref{fig:E5slow} where we treat $E_5$ as a function of $\tau_3$). The ideal $\tau_3$ gives rise to the maximum energy $E_5$ corresponding to the minimum period of twist, hence producing the optimal faster twisting somersault with orbit $\bo{L}_5(t)$. To revert to pure somersaulting motion we first find the timing $\tau_5$ (up to $\operatorname{mod} T_5$) that satisfies $E_7(\tau_5)=E_3$, meaning $\bo{L}_6(t)$ leads to $\bo{L}_7(t)=\bo{L}_3(t+c_7)$ for some phase shift $c_7$ (see Figure \ref{fig:E7slow} to uncover the desired timing $\tau_5$). We repeat this procedure to find the timing $\tau_7$ that yields $E_9(\tau_7)=E_1$, hence returning the athlete to pure somersaulting motion where $\bo{L}_9(t)=\bo{L}_1(t)=(0,l,0)^t$.

\begin{figure}[t]
\centering
\subfloat[513XD dive on $\bo{L}$-sphere.]{\includegraphics[width=7.29cm]{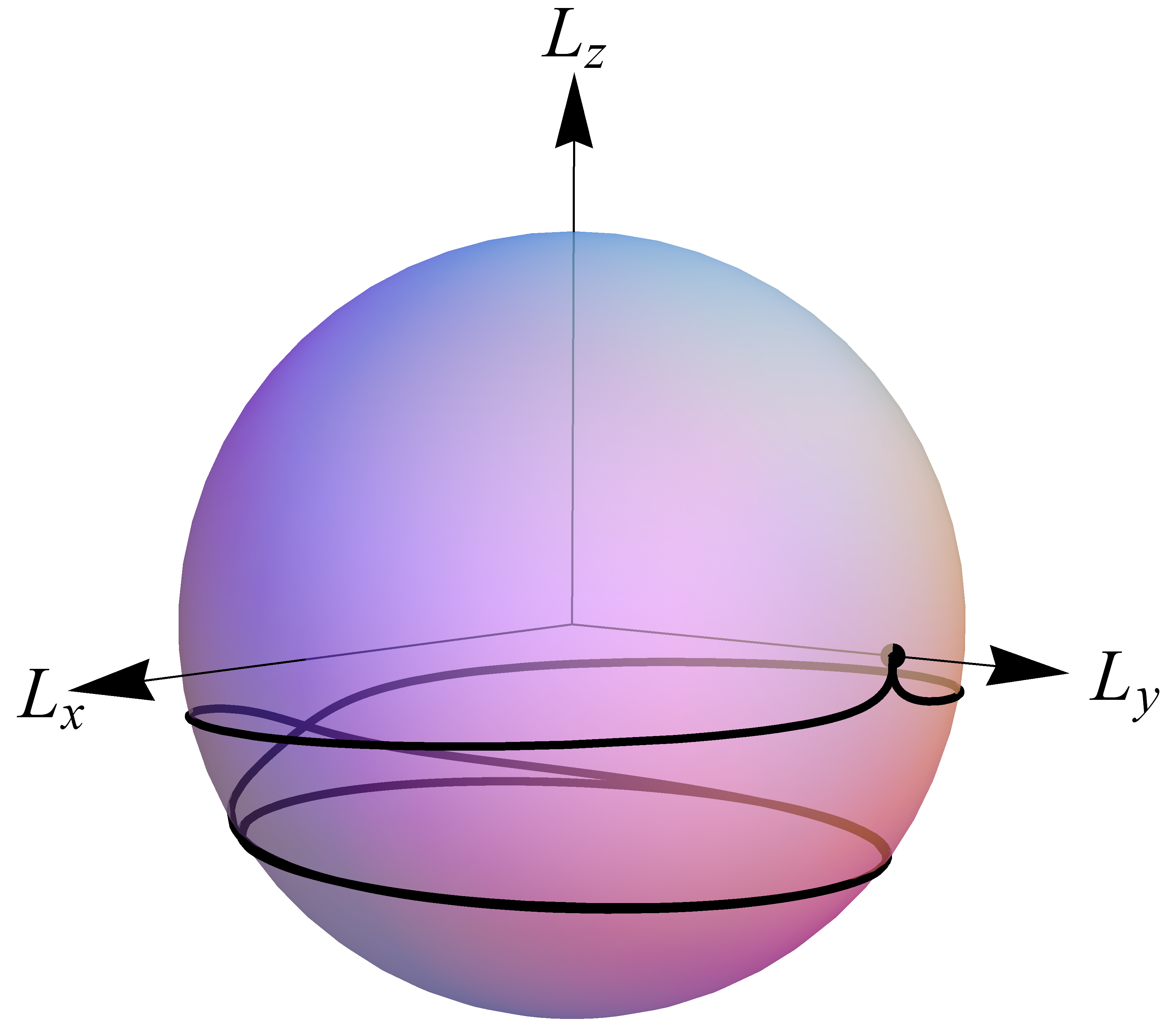}}\\
\hspace*{-15mm}\subfloat[Mercator projection with the $\operatorname{mod} 2\pi$ removed. The black points partition the trajectory into the nine different stages of the dive, and the blue shaded region gives the surface area $A$ bounded by the orbit $\bo{L}$ and equator.]{\includegraphics[width=17cm]{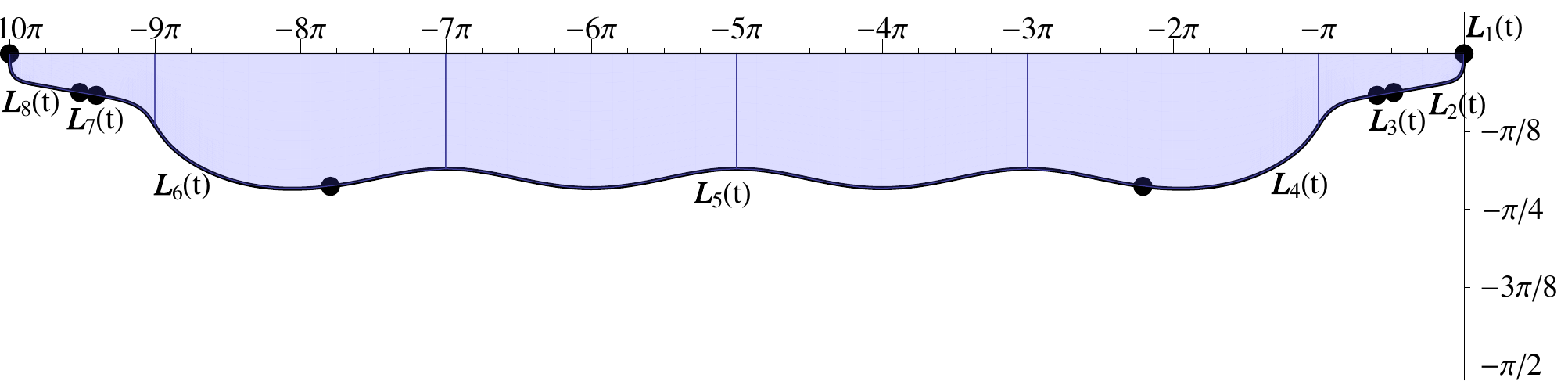}\label{fig:Lsphereallb}}
\caption{The evolution of $\bo{L}(t)$ corresponding to the 513XD dive.}\label{fig:Lsphereall}
\end{figure}
\begin{figure}[t]
\centering
\subfloat[$A_2 = 2406.62$.]{\includegraphics[width=4.7cm]{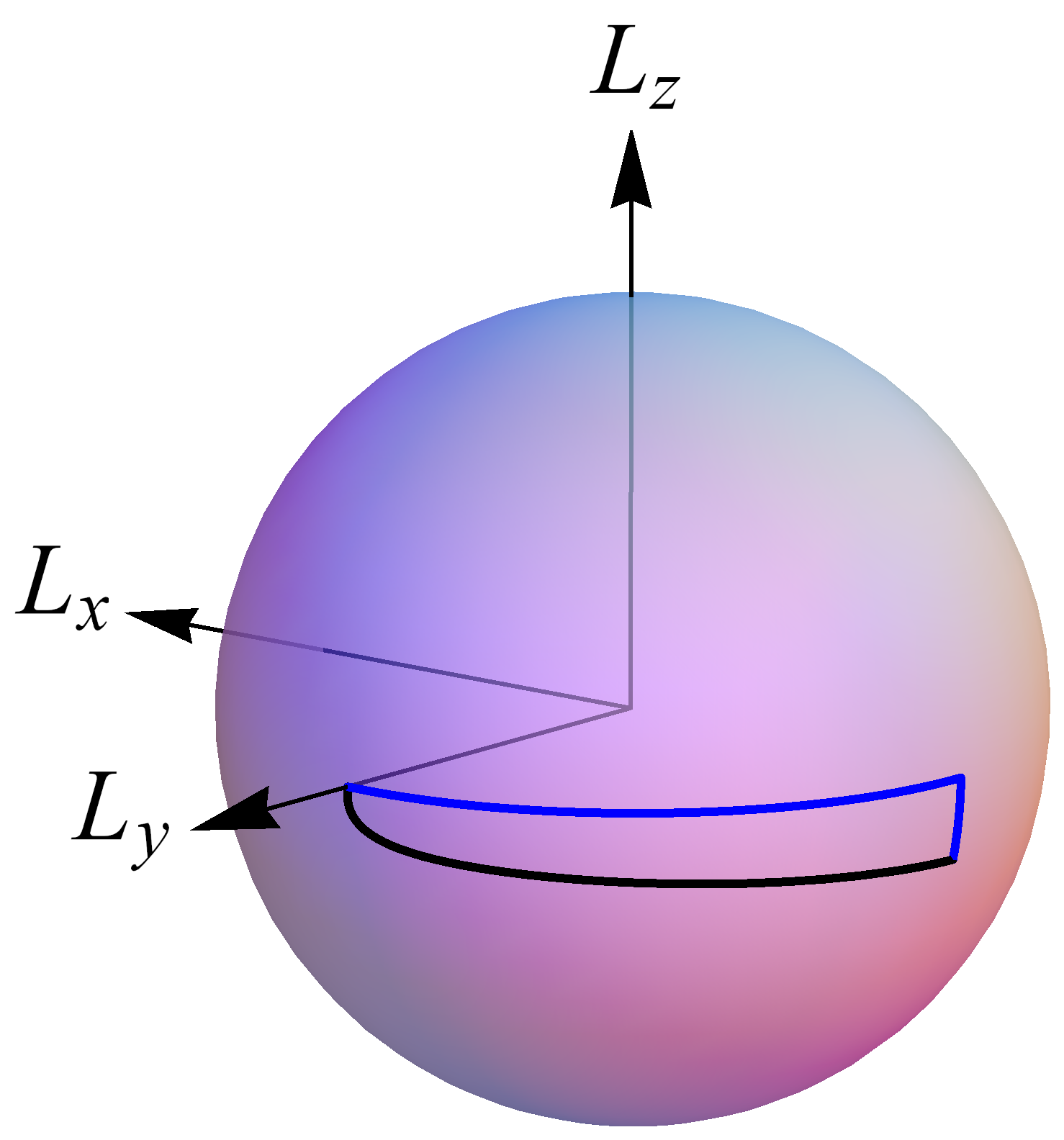}}
\hspace*{2mm}\subfloat[$A_3 = 722.037$.]{\includegraphics[width=4.7cm]{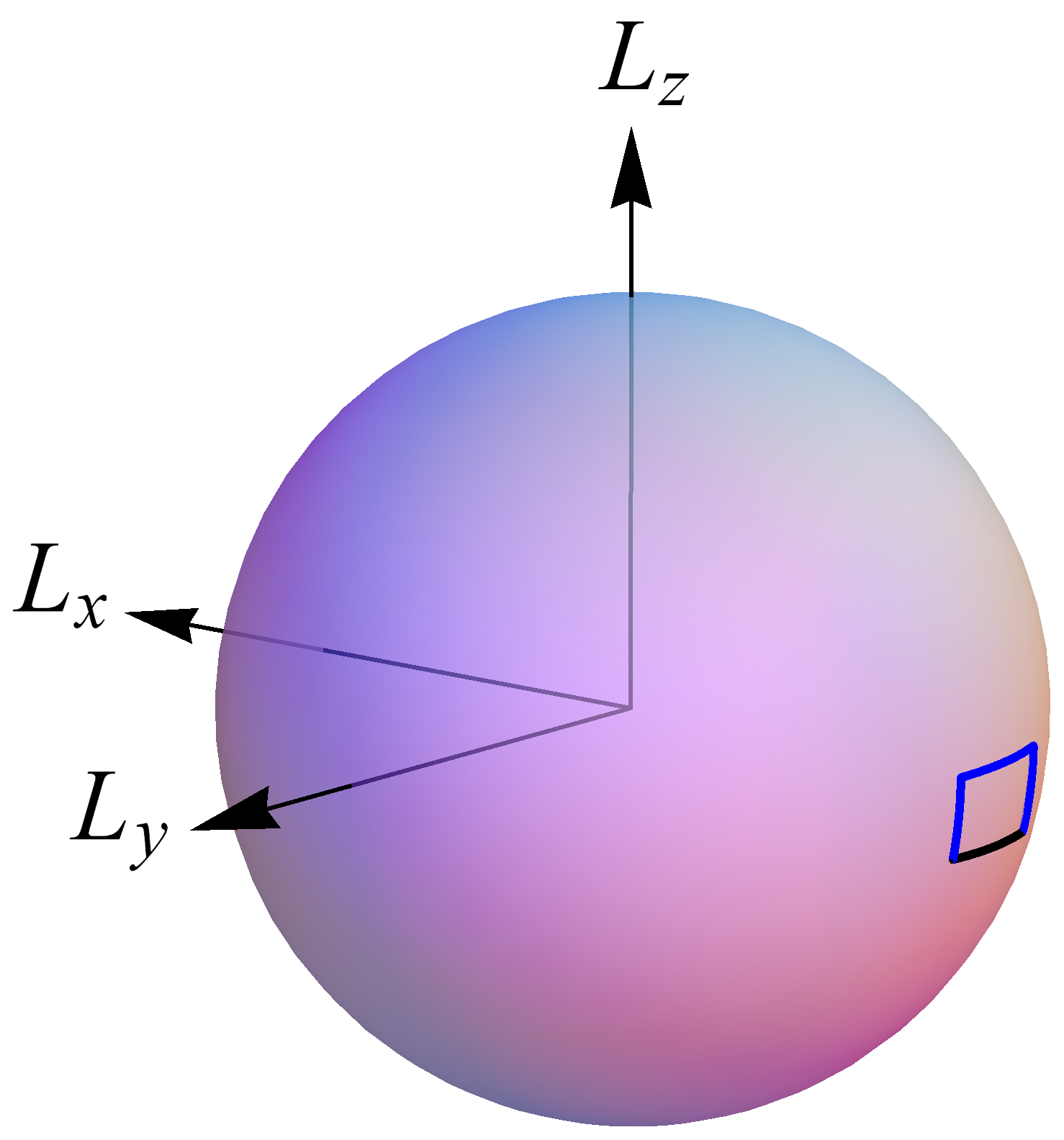}}
\hspace*{2mm}\subfloat[$A_4 = 23770.8$.]{\includegraphics[width=4.7cm]{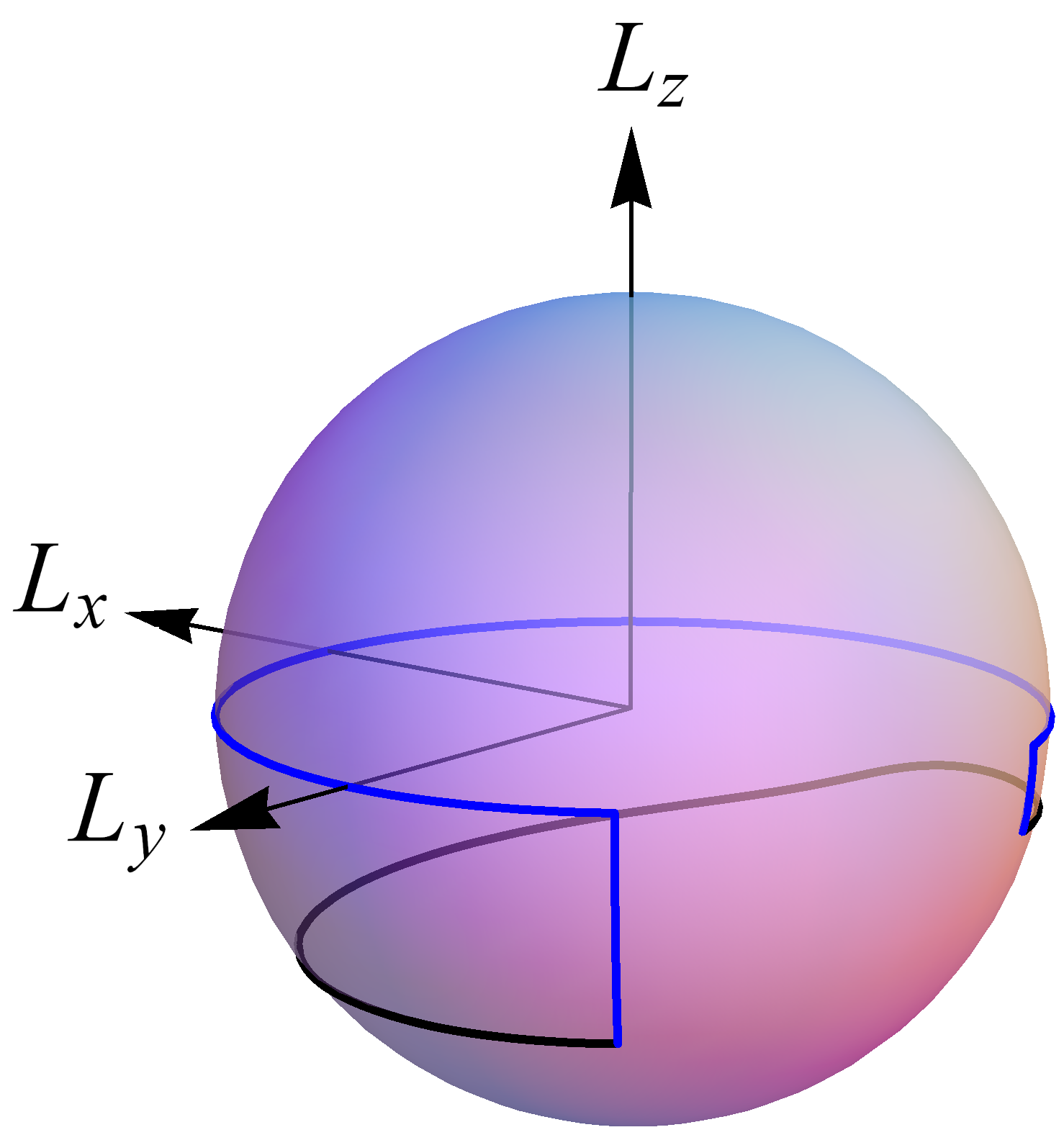}}
\caption{Due to symmetry $A_6 = A_4, A_7 = A_3$, and $A_8 = A_2$. The sub-area $A_5 = 97593.3$ can be determined from Fig.~\ref{fig:9stagekick}.}\label{fig:areas}
\end{figure}
\begin{table}[b]\centering
\begin{tabular}{|c|c|c|c|c|c|c|c|c|}
\hline
Stage  & Shape &  $E_i$			&  $T_i$ 		&  $\tau_i$ &    $A_i/l^2$ &     $2E_i \tau_i/l$	& $m_i$	& $n_i$ \\ \hline 
1 	& UU	&	$242.6$		&	$\infty$  	& $0.220$			& 0 			& 1.07	& 0.17	& 0 		\\
2 	& LU		&	$314.7$		&	$1.042$   	& $0.250$			& 0.24		& 1.57	& 0.21	& 0.24 	\\
3 	& DU	&	$460.0$		&	$0.333$ 	& $0.019$			& 0.07		& 0.17	& 0.02	& 0.06	\\
4 	& HL		&	$732.5$		&	$0.313$   	& $0.250$			& 2.38		& 3.66	& 0.20	& 0.80	\\
5 	& UD	&     $1801.2$		&	$0.115$	& $0.322$			& 9.76		& 11.60	& 0.29	& 2.79	\\
6 	& LH		&	$732.5$		&	$0.313$   	& $0.250$			& 2.38		& 3.66	& 0.20	& 0.80 	\\
7 	& DU	&	$460.0$		&	$0.333$ 	& $0.019$			& 0.07		& 0.17	& 0.02	& 0.06	\\
8 	& HU	&	$314.7$		&	$1.042$   	& $0.250$			& 0.24		& 1.57	& 0.21	& 0.24	\\
9 	& UU	&	$242.6$		&	$\infty$  	& $0.220$			& 0			& 1.07	& 0.17	& 0 		\\ \hline 
sum:	& 		& 				& 			& $0.180$			& 15.14		& 24.56	& 1.50	& 5.00	\\ \hline %
\end{tabular}
\caption{Summary of the 513XD dive with $l=100$. 
Shown are the energies $E_i$, periods $T_i$, and times $\tau_i$ per stage $i$.
For the shape changing phases 2,4,6,8 they are given by appropriate averages. 
The next columns give geometric phase and dynamics phase, and the two final columns give the number 
of somersaults and the number of twists.
In the final row the totals are shown for time, geometric phase, dynamic phase, such that $24.56 - 15.14 \approx 2\pi \, 1.5 $ gives the 
desired 1.5 somersaults, which is also given in the $m_i$ column. The final entry is the number of twists achieved.
}
\label{tab:513XD}
\end{table}
\begin{figure}[b]
\centering
\subfloat[The components of the angular momentum $\bo{L}(t)$.]{\hspace{-1.2cm}\includegraphics[width=17cm]{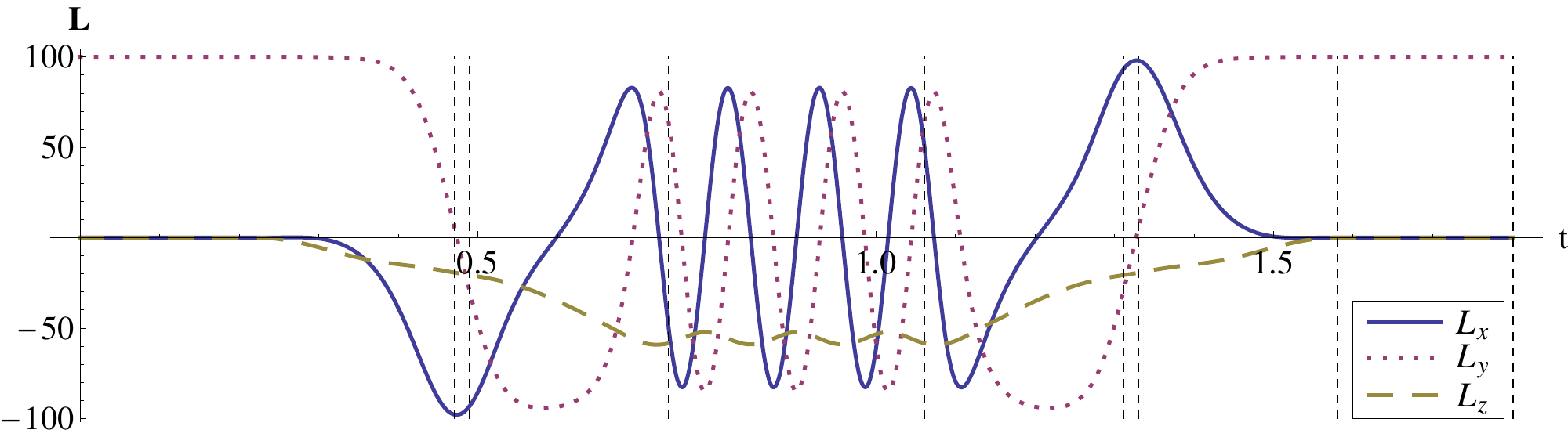}\label{newsubfig:L}}\\
\subfloat[The orientation expressed as the quaternion $q(t)$.]{\hspace{-1.2cm}\includegraphics[width=17cm]{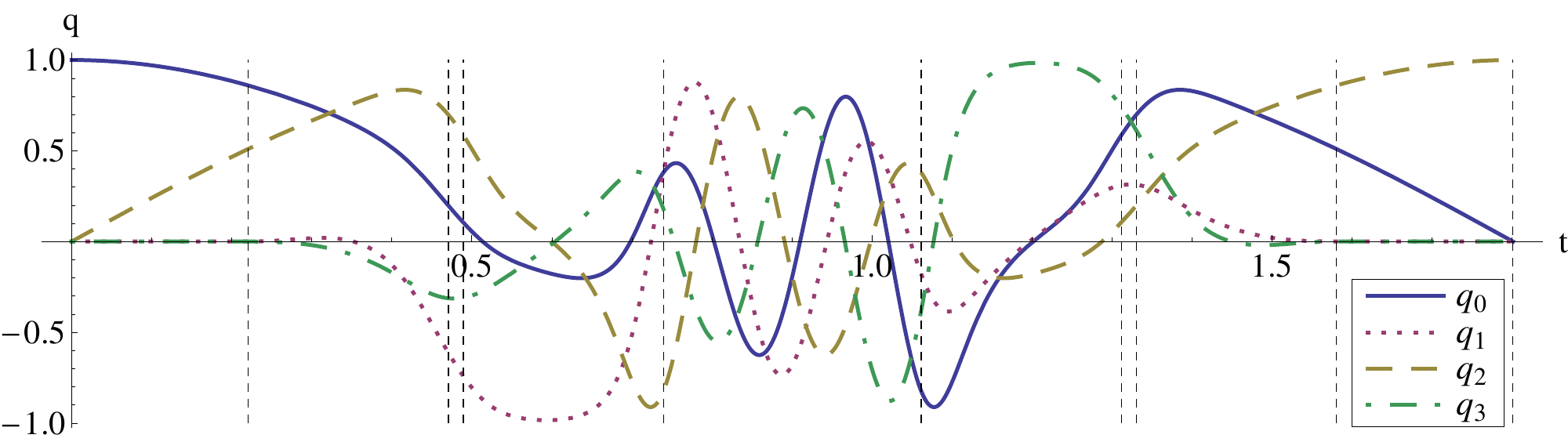}\label{newsubfig:q}}
\caption{The components of $\bo{L}(t)$ and $q(t)$ for the 513XD dive, where $q(t)$ is obtained by numerically solving \eqref{eq:eoo} given in Appendix \ref{app:eoo}. The vertical dashed lines separate the different stages of the dive.}\label{fig:Lqfullcurve}
\end{figure}

The trajectory $\bo{L}(t)$ shown in Figure \ref{fig:Lsphereall} gives rise to the faster twisting somersault consisting of 5 twists, and the 1.5 somersault constraint determines the total pure somersaulting time $\tau_1+\tau_9$. To find this, we apply Cabrera's formula \eqref{eq:cab} to the individual stages and sum up to obtain
\begin{equation}
     1.5\cdot 2\pi = -\frac{A}{l^2}+\sum_{i=1}^9 \frac{2 E_i \tau_i}{l}
\label{eq:slowkickphi}
\end{equation}
where $E_i$ is defined by 
\begin{equation}
   E_i = \frac{1}{l}\sum_{i=1}^9\int_0^{\tau_i}I^{-1}_i(t)\big(\bo{L}_i(t)-\bo{A}_i(t)\big)\cdot\bo{L}_i(t)\,dt,
\label{eq:averageE}
\end{equation}
and may be interpreted as an averaged energy during a shape changing phase. 
%
%
Since all other quantities are known we can solve for $\tau_1+\tau_9$. The surface area $A$ is illustrated by the blue shaded region in Figure \ref{fig:Lsphereallb}, which can be found by numerically computing multiple line integrals appropriately partitioned on the $\bo{L}$-sphere as shown in Figure \ref{fig:areas}. The sub-areas are computed using the line integral
$A_i = \displaystyle\oint_{\bo{C}_i} \bo{F}(\bo{x}(s))\cdot\bo{\dot{x}}(s)\,\mathrm{d}s$,
where $\bo{x}=(x,y,z)^t$, $\displaystyle \bo{F}(\bo{x})=\frac{l z}{x^2+y^2}\big(y,-x,0\big)^t$ and each loop $\bo{C}_i$ consists of a segment from the equator, one or two vertical arcs and $\bo{L}_i(t)$. Performing the computation we find that the only non-zero contribution comes from $\bo{L}_i(t)$, hence 
$A_i = \displaystyle\int_0^{\tau_i} \bo{F}\big(\bo{L}_i(t)\big)\cdot\dot{\bo{L}}_i(t)\,\mathrm{d}t$.
Evaluating the integrals yields the results shown in Figure \ref{fig:areas}, resulting in $\displaystyle A = \sum_{i=2}^8 A_i = 151392$. We give the 513XD dive summary in Table \ref{tab:513XD}, and plot the components of $\bo{L}(t)$ and $q(t)$ in Figure \ref{fig:Lqfullcurve}. We find that $L_x$ is anti-symmetric, while $L_y$ and $L_z$ are symmetric about $\mathcal{T}_\text{513XD}/2$. We solve the equations of orientation \eqref{eq:eoo} given in Appendix \ref{app:eoo} numerically and plot the result in Figure \ref{newsubfig:q}, which shows $q_0(t)=q_2(\mathcal{T}_\text{513XD}-t)$ and $q_1(t)=-q_3(\mathcal{T}_\text{513XD}-t)$. The symmetry is a result of distributing $\tau_1$ and $\tau_9$ equally. 

Table~\ref{tab:513XD} can be considered the main summary of the new 513XD dive. 
The subscripts denote the stage as usual.
The table gives the energies $E_i$, which are averaged as defined in \eqref{eq:averageE} for the shape changing phases 2, 4, 6, 8.
For the rigid phases 1, 3, 5, 7 the periods of twisting motion $T_i$ are given as defined in \eqref{eq:T},
while for the shape changing phases they are defined by dividing the time $\tau_i$ by the amount of twist $n_i$.
The next column gives the length $\tau_i$ of stage $i$. The following column gives the geometric phase per stage. 
It should be mentioned that technically speaking the geometric phase is only well defined for closed loops, 
and hence not for individual stages. The areas $A_i$ for the individual stages are not gauge invariant; however, 
in our case a natural choice of gauge is to put the reference segment fixed in the trunk of the model, 
and the numbers given for the individual stages are with respect to this choice. 
The last three columns are quantities that are derived from the earlier quantities: the dynamic phase
given by $2 E_i \tau_i / l$, the number of somersaults $m_i$ per stage (given by the difference of dynamic phase 
and geometric phase divided by $2 \pi$), and the number of twists $n_i$ per stage. 
It is interesting to compare the entries of the full dive to those of the kick model as given in Table~\ref{tab:9stage}.
Clearly energy and period for the rigid phases are very similar, which explains why the kick-model is a good
approximation. Such agreement cannot hold for $\tau_i$, since in the full model the shape changing times 
$\tau_2 = \tau_4 = \tau_6 = \tau_8 = 0.25$ must be absorbed in order to keep the total time the same.

The airborne time $\mathcal{T}_\text{513XD}=1.8$ is slightly longer than the typical 1.5-1.6 second flight time from the 10m platform, but as the angular momentum $l=100$ chosen is quite conservative we can lower $\mathcal{T}_\text{513XD}$ with a larger $l$ value. Furthermore, over $0.4$ seconds is spent in the layout position ($I_{s,y} \approx 20$) to achieve the desired somersault amount, which can be significantly reduced if the model incorporates shape changes allowing the athlete to enter pike ($I_y \approx 6$) or tuck ($I_y \approx 4$) positions. This leads us to conclude that real world athletes can in principle execute the 513XD dive, and this would revolutionise the sport of diving if successfully performed in competition. 

\section{From Pure Somersault to Pure Twist}\label{sec:som2twi}

\begin{figure}[b]
\centering
\hspace*{-2.2mm}\subfloat[Pure somersaulting state.]{\includegraphics[width=7.25cm]{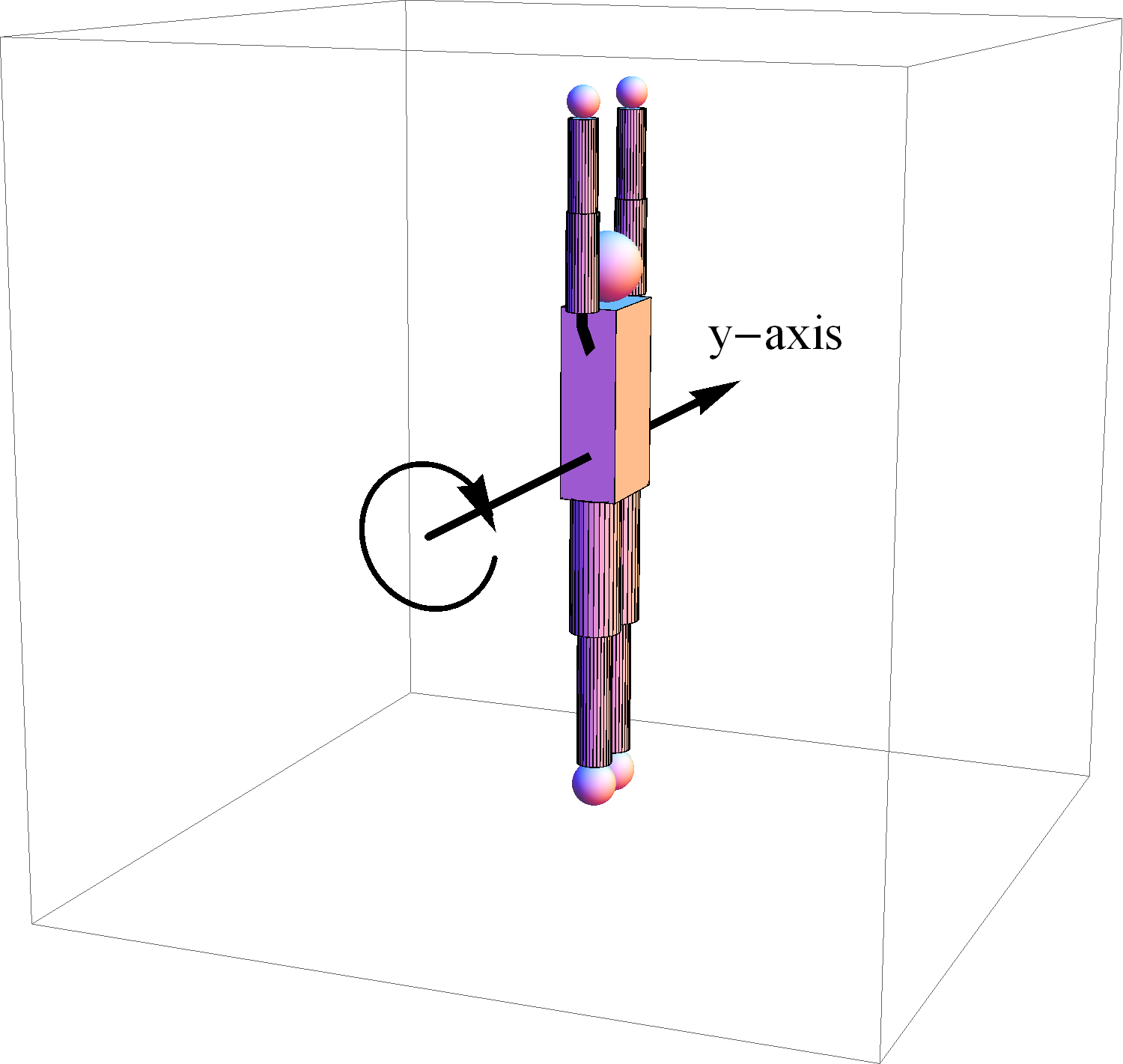}\label{subfig:puresom}}
\hspace*{5mm}\subfloat[Pure twisting state.]{\includegraphics[width=7.25cm]{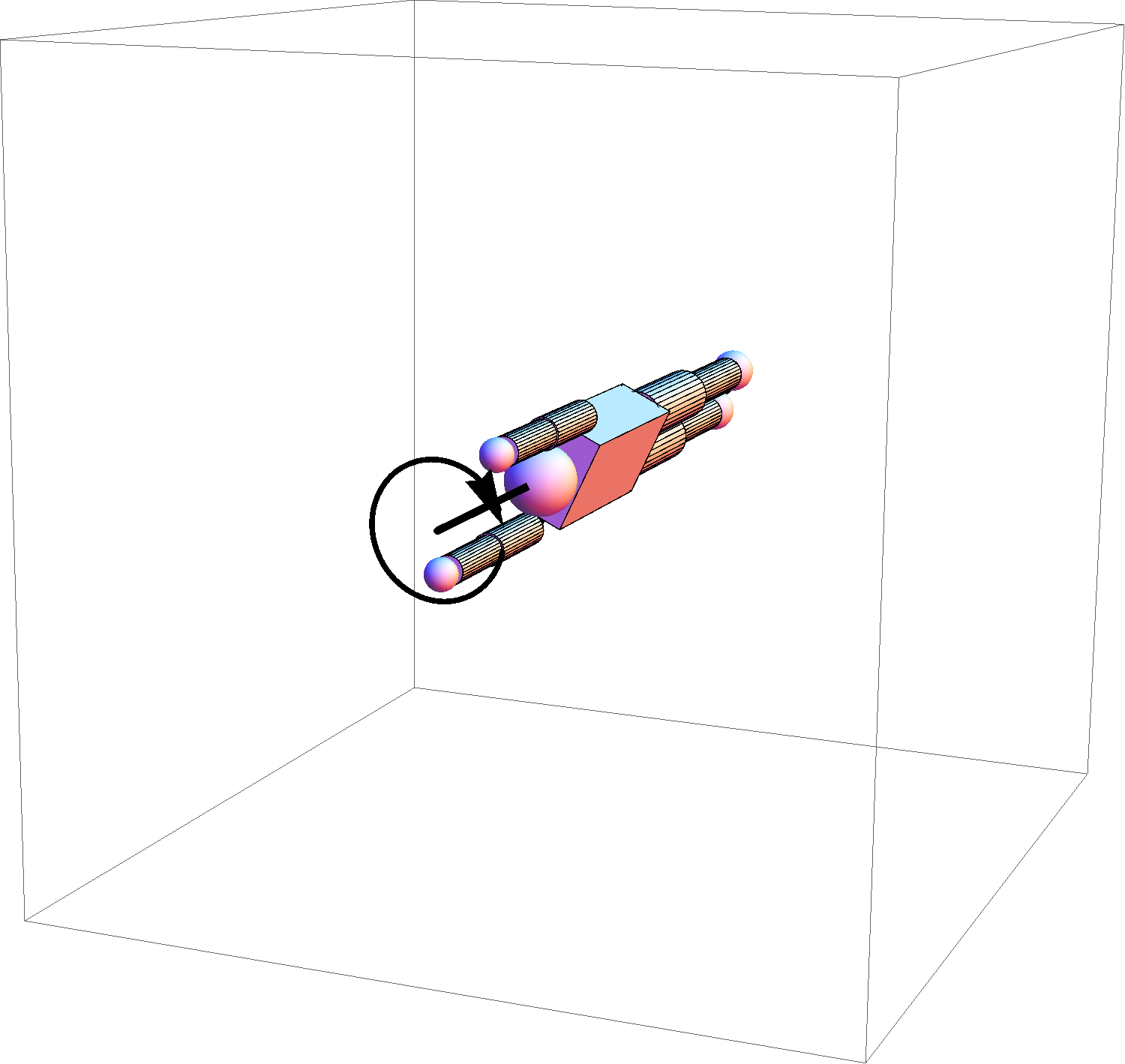}\label{subfig:puretwi}}
\caption{The pure somersaulting and pure twisting states.}
\end{figure}

The 513XD dive uses an extra pair of shape changes in addition to the 5-stage dive to enter and exit the faster twisting somersaulting motion. For impulsive shape changes we found that the optimal fast-kick timing is after a half twist, so there is no reason why this procedure cannot be repeated to further speed up twist in the twisting somersault. We now show how an athlete taking off in pure somersaulting motion (Figure \ref{subfig:puresom}) can use a sequence of impulsive shape changes to enter a state of pure twist (Figure \ref{subfig:puretwi}),
assuming a sufficiently large overall time. 
This question was first discussed in Yeadon's thesis \cite{YeadonThesis}. 

\begin{figure}[t]
\includegraphics[width=7.4cm]{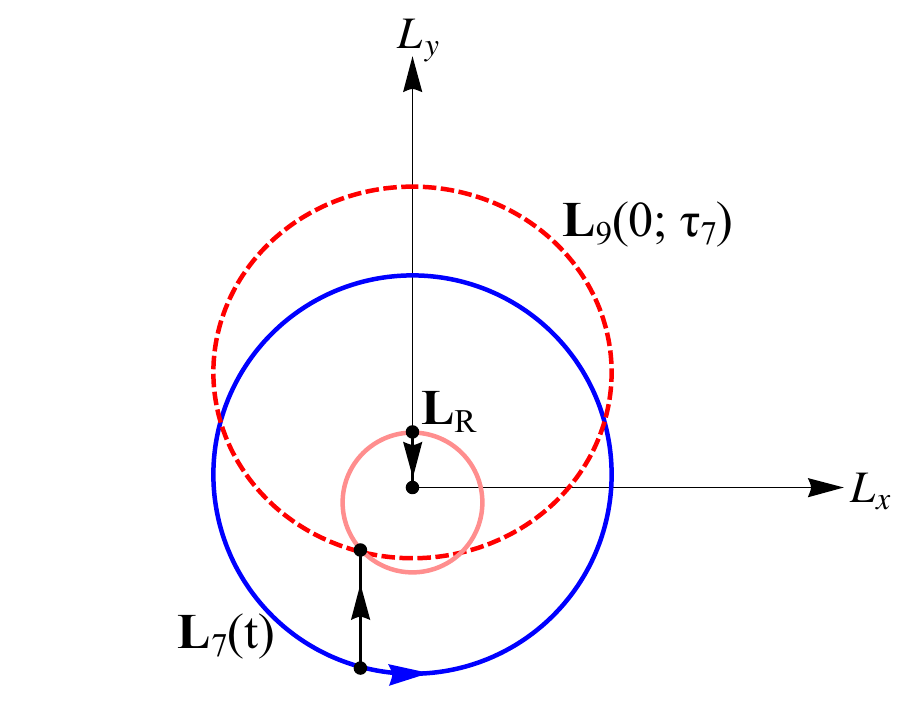}
\caption{The blue outer orbit corresponds to $\bo{L}_7(t)$, the red inner orbit represents $\bo{L}_R(t)$ and the red dashed orbit $\bo{L}_9(0;\tau_7)$ is the family of possible initial conditions generated by HL fast-kick from $\bo{L}_7(\tau_7)$. The black vertical lines show the impulsive shape change transition from $\bo{L}_7(t)$ to $\bo{L}_R(t)$ orbit and from $\bo{L}_R(t)$ to $\bo{l}_t$.}
\label{fig:twistorbit}
\end{figure}
\begin{figure}[b]
\includegraphics[width=7.4cm]{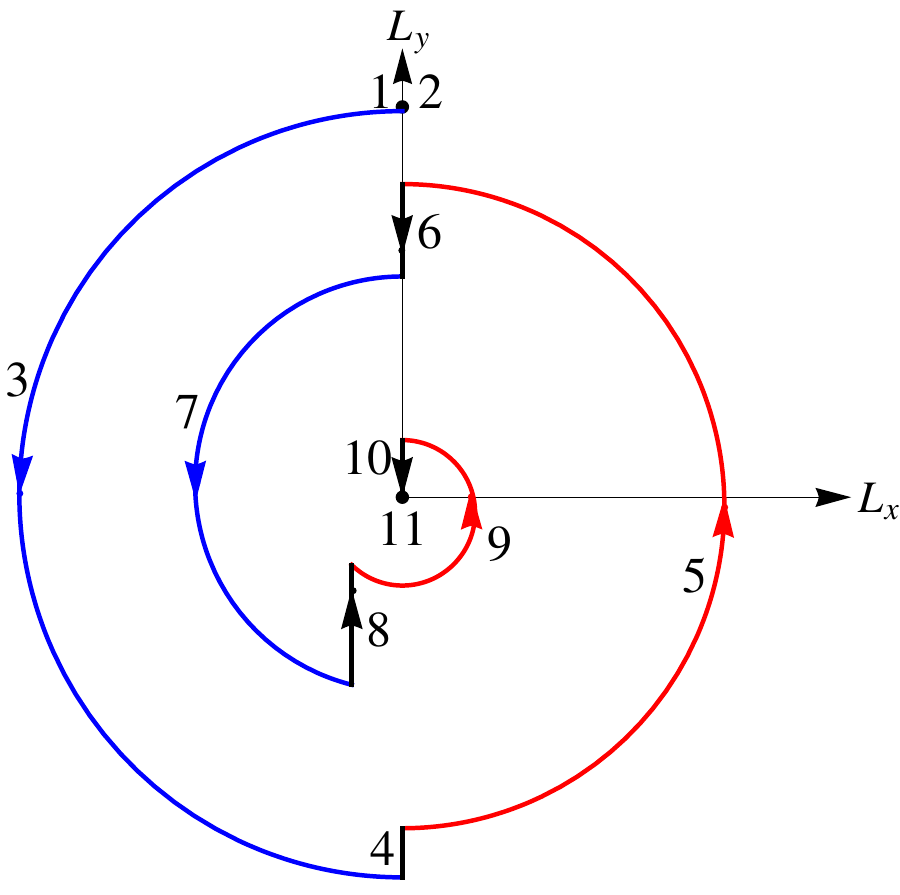}
\caption{$\bo{L}$-sphere projection onto the $xy$-plane for the pure somersault to pure twist dive. }
\label{fig:som2twi}
\end{figure}

The pure twisting motion corresponds to the steady rotation $\bo{l}_t = (0,0,-l)^t$ with period 
\begin{equation}
T_t=2\pi I_{s,z}/l=6.2553/l.\end{equation}
The twists are in the counterclockwise direction and the energy 
\begin{equation}
E_t=l^2 I_{s,y}^{-1}/2 = 0.5022l^2
\end{equation}
is maximal for the UU shape. 

The athlete begins by performing fast-kicks after half twist intervals to speed up the twisting motion, but beyond a certain point (after stage 7 with $\bo{L}_7(t)$) no further improvement is made and the athlete needs to adopt a new strategy. Suppose $\bo{L}_R(t)$ corresponds to the rigid body orbit with shape UH that contains the point $\bo{L}_R=\mathcal{R}_x(\mathcal{X})\bo{l}_t$, then we know a UH fast-kick at this point will lead to pure twisting motion. Similarly, let $\bo{L}_L(t)$ correspond to the symmetric case with point $\bo{L}_L=\mathcal{R}_x(-\mathcal{X})\bo{l}_t$. Our goal is now to find (if possible) either a HL or LH fast-kick timing that brings the athlete into $\bo{L}_R(t)$ or $\bo{L}_L(t)$ orbit, respectively. We show in Figure \ref{fig:twistorbit} that there are two points of intersection between $\bo{L}_9(0;\tau_7)$ and $\bo{L}_R(t)$, where the earlier timing occurs at $\tau_7=3.4946/l$. The athlete then remains on the $\bo{L}_R(t)$ orbit for $\tau_9=4.0735/l$ to reach the designated point $\bo{L}_R$ before executing the UH fast-kick to enter a state of pure twist. The complete transition from pure somersault to pure twist is given in Table \ref{tab:som2twi}. Although the realistic analogue of this cannot be performed in platform diving, it nevertheless has interesting applications. For example, it may be possible to achieve momentary pure twist in sports that have larger airborne times, such as high diving and aerial skiing. Also, the conversion from pure somersault to pure twist (and vice-versa) has applications in space manoeuvrability where airborne time is not a factor.
\begin{table}[h]\centering
\begin{tabular}{cccc}
\hline
Stage & Energy $E_i$  & Time spent $\tau_i$	& Initial condition $\bo{L}_i(0)$\\
Shape & Period $T_i$  & Twist rate			& Orbit $\bo{L}_i(t)$\\\hline\hline
1 		& $0.0243l^2$   & $-$							& $\bo{l}$\\
UU		&	$\infty$			& $0$							& $(0,l,0)^t$\\\hline\hline
3			& $0.0452l^2$		& $T_3/2$					& $\mathcal{R}_x(-\mathcal{X})\bo{l}$\\
DU		& $33.9610/l$		& $0.18501 l$			& $R_p \bo{\mathcal{L}}(t; E_3, J_t, T_3/4)$\\\hline\hline
5 		& $0.1835l^2$ 	& $T_5/2$					& $-\mathcal{R}_x(\mathcal{X}+\mathcal{Y}+2p)\bo{l}$\\
UD		&	$11.3854/l$		& $0.55187 l$			& $R_p^{-1} \bo{\mathcal{L}}(t; E_5, J_t, 3T_5/4)$\\\hline\hline
7			& $0.3786l^2$		& $3.4946/l$			& $\mathcal{R}_x(-\mathcal{X}-2\mathcal{Y}-4p)\bo{l}$\\
DU		& $7.5855/l$		& $0.8283 l$			& $R_p \bo{\mathcal{L}}(t; E_7, J_t, T_7/4)$\\\hline\hline
9 		& $0.4996l^2$ 	& $4.0735/l$			& $(-0.130, -0.174, -0.976)^t$\\
UD		&	$6.5410/l$		& $0.9606 l$			& $R_p^{-1} \bo{\mathcal{L}}(t; E_9, J_t, 4.1028)$\\\hline\hline
11 		& $E_t=0.5022l^2$& $-$						& $\bo{l}_t$\\
UU		&	$T_t=6.2553/l$& $1.0045 l$			& $(0,0,-l)^t$\\\hline\hline
\hline
\end{tabular}
\caption{Sequence of impulsive shape changes to transition from a state of pure somersault (stage 1) to a state of pure twist (stage 11).}
\label{tab:som2twi}
\end{table}

\section{Acknowledgement}

This research was supported in part by the Australian Research Council through the Linkage Grant LP100200245 ``Bodies in Space'' 
in collaboration with the New South Wales Institute of Sports.

\appendix
\section{Model Parameters}\label{app:modelparameters} 
\begin{table}[h]\centering\begin{tabular}{| c | l | c | l l l|}\hline
segment & \multicolumn{1}{c|}{partitions}				& \multicolumn{1}{c|}{mass (kg)} & \multicolumn{3}{c|}{geometry (cm)}\\\hline
\multirow{8}{*}{$B_b$} & head						& 5.575			& sphere:		&$r=11$		&\\
			& torso 					& 32.400\phantom{*}		& cuboid: 	&\multicolumn{2}{l|}{$18\times30\times60$}\\		
			& left thigh			& 8.650			& cylinder:	&$r=8$		&$h=43$\\
			& right thigh			& 8.650			& cylinder:	&$r=8$		&$h=43$\\
			& left lower leg	& 4.086			& cylinder:	&$r=5.5$	&$h=43$\\
			& right lower leg	& 4.086			& cylinder:	&$r=5.5$	&$h=43$\\
			& left foot				& 1.436			& sphere:		&$r=7$		&\\
			& right foot			& 1.436			& sphere:		&$r=7$		&\\\hline
			& left upper arm	& 2.356			& cylinder:	&$r=5$		&$h=30$\\
$B_l$	& left forearm		& 1.781			& cylinder:	&$r=4.5$	&$h=28$\\
			& left hand				& 0.523			& sphere:		&$r=5$		&\\\hline
			& right upper arm	& 2.356			& cylinder:	&$r=5$		&$h=30$\\
$B_r$	& right forearm		& 1.781			& cylinder:	&$r=4.5$	&$h=28$\\
			& right hand			& 0.523			& sphere:		& $r=5$&\\\hline
\end{tabular}
\caption{Frohlich's twelve segment model of a male athlete that is 1.82 m in height and weighs 75.639 kg.}
\label{tab:dimensions}
\end{table}
We combine multiple body parts of Table \ref{tab:dimensions} to produce the three segments denoted by $B_i$ for $i\in\{b,l,r\}$, which represent the body, left arm and right arm, respectively. 
The numerical values of the mass and tensor of inertia of segments in our model are
\begin{align}
m_b &= 66.319 &  \tilde{I}_b &= \diag{(14.204, 13.867, 0.612)}\nonumber\\
m_l&=m_r=4.660 & \tilde{I}_l &= \tilde{I}_r = \diag{(0.176, 0.176, 0.005)}.\nonumber
\end{align}
The collection of $\{\bo{\tilde{J}}_i^j\}$ that specify the geometry of our model are
\begin{align}
\bo{\tilde{J}}_b^l &= (0, 0.2, 0.5196)^t & \bo{\tilde{J}}_b^r &= (0, -0.2, 0.5196)^t & \bo{\tilde{J}}_l^b&=\bo{\tilde{J}}_r^b=(0, 0, 0.3647)^t.\nonumber
\end{align}
In the layout position the athlete has shape $(\alpha_l,\alpha_r)=(\pi,\pi)$ and the tensor of inertia is
\begin{equation}
I_s=\diag(21.3188, 20.6091, 0.9956).\nonumber
\end{equation}
In the twist position the shape is either $(\alpha_l,\alpha_r)=(0,\pi)$ or $(\alpha_l,\alpha_r)=(\pi,0)$, which produces the same diagonalised tensor of inertia
\begin{equation}
J_t=\diag(18.3745, 17.6925, 0.9679).\nonumber
\end{equation}

\section{Components of $I$ and $\bo{A}$}\label{app:components} 
Evaluating \eqref{eq:I} with the abduction-adduction plane of motion restriction simplifies the tensor of inertia $I$ to the form
$I = \displaystyle\left(\begin{array}{ccc}
I_{xx} & 0      & 0\\
0      & I_{yy} & I_{yz}\\
0      & I_{yz} & I_{zz}
\end{array}\right)$.
Explicitly, the components are 
\begin{align}
I_{xx} &= a_0 - 2a_1\cos_{{}_+}{\!(\alpha_l,\alpha_r)} + 2a_2\sin_{{}_+}{\!(\alpha_l,\alpha_r)}- 2a_3\cos{(\alpha_l + \alpha_r)}\nonumber\\
I_{yy} &= a_5 - 2a_1\cos_{{}_+}{\!(\alpha_l,\alpha_r)} + a_4\cos_{{}_+}{\!(2\alpha_l,2\alpha_r)}- 2a_3\cos{\alpha_l}\cos{\alpha_r}\nonumber\\
I_{zz} &= a_6 + 2a_2\sin_{{}_+}{\!(\alpha_l,\alpha_r)} - a_4\cos_{{}_+}{\!(2 \alpha_l,2\alpha_r)}+ 2a_3\sin{\alpha_l} \sin{\alpha_r}\nonumber\\
I_{yz} &= a_2\cos_{{}_-}{\!(\alpha_l,\alpha_r)} - a_1\sin_{{}_-}{\!(\alpha_l,\alpha_r)} + a_4\sin_{{}_-}{\!(2 \alpha_l,2\alpha_r)} - a_3\sin{(\alpha_l-\alpha_r)},\nonumber
\end{align}
where
\begin{align}
\cos_{{}_+}{\!(\alpha_l,\alpha_r)} &= \cos{\alpha_l}+\cos{\alpha_r} & \sin_{{}_+}{\!(\alpha_l,\alpha_r)} &= \sin{\alpha_l}+\sin{\alpha_r}\nonumber\\
\cos_{{}_-}{\!(\alpha_l,\alpha_r)} &= \cos{\alpha_l}-\cos{\alpha_r} & \sin_{{}_-}{\!(\alpha_l,\alpha_r)} &= \sin{\alpha_l}-\sin{\alpha_r}.\nonumber
\end{align}
Similarly, \eqref{eq:A} simplifies to $\bo{A}=\big(A_l \dot{\alpha}_l+A_r \dot{\alpha}_r,0,0\big)^t$ where
\begin{align}
A_l &= a_7 - a_1 \cos{\alpha_l}+ a_2 \sin{\alpha_l} - a_3 \cos{(\alpha_l+\alpha_r)}\nonumber\\
A_r &= -a_7 + a_1 \cos{\alpha_r}- a_2 \sin{\alpha_r} + a_3 \cos{(\alpha_l+\alpha_r)}.\nonumber
\end{align}
The constants $a_0, a_1,\dots,a_7$ are determined by the collection of $\left\{m_i, \tilde{I}_i,\bo{\tilde{J}}_i^j\right\}$ and are
\begin{align}
a_0 &= 18.298 & a_1 &= 0.774 & a_2 &= 0.340 & a_3 &= 0.038\nonumber\\
a_4 &= 0.376  & a_5 &= 16.836& a_6 &= 1.748 & a_7 &= 0.758.\nonumber
\end{align}

\section{Equations of Orientation}\label{app:eoo}
The orientation can be tracked from the solution of the equations of motion \eqref{eq:eom}. We will represent the orientation with unit quaternions as they provide an elegant form of encoding the angle-vector information. Consider a clockwise rotation of $\theta$ about the unit vector $\bo{u}$, which can be presented with the unit quaternion $q=q_0+\bo{q}=\cos(\theta/2)+\bo{u}\sin{\theta/2}$, where the vector $\bo{q}=q_1\bo{i}+q_2\bo{j}+q_3\bo{k}$ specifies the imaginary parts. To rotate an arbitrary vector $\bo{v}$ by the quaternion $q$, we first treat the vector as a pure quaternion expressed as $v=0+\bo{v}$ and then apply the transformation $p=qv\bar{q}$, where $\bar{q}=q_0-\bo{q}$ is the quaternion conjugate. The result is a pure quaternion $p=0+2(\bo{v}\cdot\bo{q})\bo{q}+(q_0^2-\bo{q}\cdot\bo{q})\bo{v}-2q_0\bo{v}\times\bo{q}$, which is linear in $\bo{v}$ and can therefore be rearranged to obtain the vector
\begin{equation}
\bo{p}=\left[2(\bo{q}\bo{q}^t+q_0 \bo{q})+(q_0^2-\bo{q}\cdot\bo{q})\mathbbm{1}\right]\bo{v}.
\end{equation}
Now the coefficient of $\bo{v}$ is precisely the rotation matrix $R$, so substituting it in $\hat{\bo{\Omega}}=R^t\dot{R}$ and removing the hat operator gives
\begin{equation}
\bo{\Omega}=2\left(\begin{array}{cccc}
-q_1 &  q_0 &  q_3 & -q_2\\
-q_2 & -q_3 &  q_0 &  q_1\\
-q_3 &  q_2 & -q_1 &  q_0
\end{array}\right)\left(\begin{array}{c}
\dot{q}_0\\\dot{q}_1\\\dot{q}_2\\\dot{q}_3\end{array}\right),\label{eq:qderive}
\end{equation}
where $\bo{\Omega}$ is a known vector obtained from solving the equations of motion \eqref{eq:eom}. As $q$ is a unit quaternion we can incorporate the constraint $q_0\dot{q}_0+q_1\dot{q}_1+q_2\dot{q}_2+q_3\dot{q}_3=0$ with \eqref{eq:qderive} to derive the equations of orientation 
\begin{equation}
\dot{q} = \frac{1}{2}\left(\begin{array}{cc}
					0 & -\bo{\Omega}^t\\
\bo{\Omega} & -\hat{\bo{\Omega}}\end{array}\right) q.\label{eq:eoo}
\end{equation}
Together with \eqref{eq:eom} and \eqref{eq:eoo}, a complete description of the dynamics for a system of coupled rigid bodies can be given.
\bibliographystyle{siam}
\bibliography{513XD}
\end{document}